\documentclass[journal,10pt]{IEEEtran}

\usepackage{graphicx}
\usepackage[colorlinks,linkcolor=black,anchorcolor=black,citecolor=black]{hyperref}
\usepackage{lineno,hyperref}
\modulolinenumbers[5]
\usepackage[cmex10]{amsmath}
\usepackage{amsfonts}

\usepackage{color}
\usepackage{algorithm}
\usepackage{algorithmic}
\usepackage{array}
\usepackage{multirow}
\usepackage{booktabs}
\usepackage{subfigure}
\usepackage{bm}
\usepackage{graphicx}
\usepackage{cite}

\allowdisplaybreaks[4]

\usepackage{amsmath}
\usepackage{underscore}
\usepackage{stfloats}
\usepackage{float}
\usepackage{lipsum}
\usepackage{mathrsfs}
\usepackage{amssymb}
\usepackage{textcomp}
\makeatletter
\def\subsubsection{\@startsection{subsubsection}{3}{\parindent}{0.5ex plus 1ex minus 0.1ex}{0pt}{\normalfont\normalsize}}

\newenvironment{breakablealgorithm}
{
	\begin{center}
		\refstepcounter{algorithm}
		\hrule height.8pt depth0pt \kern2pt
		\renewcommand{\caption}[2][\relax]{
			{\raggedright\textbf{\ALG@name~\thealgorithm} ##2\par}%
			\ifx\relax##1\relax 
			\addcontentsline{loa}{algorithm}{\protect\numberline{\thealgorithm}##2}%
			\else 
			\addcontentsline{loa}{algorithm}{\protect\numberline{\thealgorithm}##1}%
			\fi
			\kern2pt\hrule\kern2pt
		}
	}{
		\kern2pt\hrule\relax
	\end{center}
}

\begin{document}
\title{Hybrid Hierarchical DRL Enabled Resource Allocation for Secure Transmission in Multi-IRS-Assisted Sensing-Enhanced Spectrum Sharing Networks}

\author{Lingyi Wang,
Wei Wu,~\IEEEmembership{Member,~IEEE,}
Fuhui Zhou,~\IEEEmembership{Senior Member,~IEEE,}
Qihui Wu,~\IEEEmembership{Senior Member,~IEEE,}\newline
Octavia A. Dobre,~\IEEEmembership{Fellow,~IEEE,}
and Tony Q.S. Quek,~\IEEEmembership{Fellow,~IEEE}

\thanks{This work was supported in part by the National Key Research and Development Program of China under Grant 2020YFB1807602;
in part by the National Natural Science Foundation of China under Grant 62271267, Grant 62222107, Grant 62071223 and Grant 62031012;
in part by the Zhejiang Lab Open Research Project under Grant K2022PD0AB09;
in part by the Natural Sciences and Engineering Research Council of Canada (NSERC) through its Discovery program;
and in part by the National Research Foundation, Singapore and Infocomm Media Development Authority under its Future Communications Research \& Development Programme.}
\thanks{Lingyi Wang is with the College of Science,
Nanjing University of Posts and Telecommunications, Nanjing, 210003, China
(e-mail: lingyiwang@njupt.edu.cn).}
\thanks{
Wei Wu is with the College of Communication and Information Engineering,
Nanjing University of Posts and Telecommunications, Nanjing, 210003, China
(e-mail: weiwu@njupt.edu.cn).}
\thanks{Fuhui Zhou and Qihui Wu are with the College of Electronic and Information Engineering, 
Nanjing University of Aeronautics and Astronautics, Nanjing, 210000, China
(e-mail: zhoufuhui@ieee.org; wuqihui2014@sina.com).}
\thanks{Octavia A. Dobre is with the Faculty of Engineering and Applied Science, Memorial University, St. John's, NL A1B 3X5, Canada 
(e-mail: odobre@mun.ca).}
\thanks{T. Q. S. Quek is with the Singapore University of Technology and Design, 487372, Singapore (e-mail: tonyquek@sutd.edu.sg).}
}

\maketitle

\begin{abstract}
  Secure communications are of paramount importance in spectrum sharing networks due to the allocation and sharing characteristics of spectrum resources.
  To further explore the potential of intelligent reflective surfaces (IRSs) in enhancing spectrum sharing and secure transmission performance, a multiple intelligent reflection surface (multi-IRS)-assisted sensing-enhanced wideband spectrum sharing network is investigated by considering physical layer security techniques. 
  An intelligent resource allocation scheme based on double deep Q networks (D3QN) algorithm and soft Actor-Critic (SAC) algorithm is proposed to maximize the secure transmission rate of the secondary network 
  by jointly optimizing IRS pairings, subchannel assignment, transmit beamforming of the secondary base station, reflection coefficients of IRSs and the sensing time.  
  To tackle the sparse reward problem caused by a significant amount of reflection elements of multiple IRSs, the method of hierarchical reinforcement learning is exploited.
  An alternative optimization (AO)-based conventional mathematical scheme is introduced to verify the computational complexity advantage of our proposed intelligent scheme.
  Simulation results demonstrate the efficiency of our proposed intelligent scheme as well as the superiority of multi-IRS design in enhancing secrecy rate and spectrum utilization. 
  It is shown that inappropriate deployment of IRSs can reduce the security performance with the presence of multiple eavesdroppers (Eves), and the arrangement of IRSs deserves further consideration. 
\end{abstract}  
  

\IEEEpeerreviewmaketitle
\section{Introduction}
\IEEEPARstart{D}{ue} to the demand for ultra-massive connectivity, ultra-high data rate and ultra-low transmission latency,
spectrum resources in the sixth-generation (6G) wireless communication are increasingly scarce.
Therefore, the development of spectrum-efficient, energy-efficient and reliable wireless communication technologies is of crucial importance for future wireless communication networks.
Both orthogonal frequency division multiplexing (OFDM) and cognitive radio (CR) have received widespread attention for high spectral efficiency.
By dividing the channel into multiple independent orthogonal subchannels, OFDM can avoid inter-subchannel interference and provide great communication quality\cite{wu2022intelligent,wang2023intelligent,aoudia2021end,9390405}.
Moreover, due to the dynamic spectrum access characteristics, CR can improve spectrum efficiency\cite{wu2023joint,wu2021resource,zhu2022dynamic}.

Besides spectrum utilization issues, secure transmission also faces serious challenges due to the characteristics of open and shared wireless spectrum.
Physical layer security (PLS) technologies have received extensive attention due to the low cost overhead and convenient implementation\cite{9237455,9201094,9844767},
which can leverage the inherent features of wireless channels to improve the signal quality of legitimate users while suppressing eavesdropping.
Artificial noise\cite{9350170}, cooperative jamming\cite{9154423} and cooperative relaying\cite{9244625} are common PLS techniques and have been widely investigated.
However, the transmitting jamming or artificial noise, and deploying relays result in additional considerable energy consumption\cite{9483903}.
Thus, it is meaningful to find new secure communication solutions to meet the low-energy and high-efficiency requirements of future 6G communication networks.

Intelligent reflection surface (IRS) has been identified as a disruptive technology for the future generation of wireless communication networks, which is capable of reshaping the signal propagation in a low-cost and low-energy manner \cite{9483903,zheng2022survey,cui2019secure}.
Specifically, the IRS is a reflection array consisting of a substantial number of passive reflection elements and a minor central control unit that can dynamically and adaptively adjust the phase of the elements in response to real-time channel information \cite{gong2020toward}.
Compared to traditional active transceivers/relays, IRS is more energy-efficient since it merely reflects signals with passive elements instead of injecting power for amplification. 
Moreover, due to programmability and reconfigurability, the reflective elements of the IRS can be adjusted adaptively to reshape the radio propagation environment and improve the transmission performance. 
For example, the signal reflected by the IRS can form enhanced beamforming with the wireless signal from the base station to enhance the desired power gain at legitimate users \cite{wu2022intelligent}. 
Furthermore, with the presence of eavesdroppers (Eves), the reflected signal can be applied to suppress the power of the signal received at the Eve, achieving a better secure performance \cite{wang2023intelligent}.
However, due to the limitation of reflection array adjustment, it is difficult for a single IRS to meet the communication requirements of large-scale users in an ultra-wide range \cite{9133435}.
Hence, to further improve system reliability and enlarge the coverage, multiple IRSs are employed for high spectral efficiency and secure communications.

However, resource allocation problems in IRS-assisted spectrum sharing networks are always multivariate coupled. 
Moreover, in the case of multiple IRSs, reflection array elements, selection of multiple reflection path, pairing between IRS and users, subchannel assignment, primary and secondary network interference management, etc., become more complex since the number of IRS increases,
which undoubtedly increases the difficulty of resource allocation in multi-IRS-assisted spectrum sharing networks. 
Additionally, traditional numerical optimization based mathematical methods cannot adapt to the complex communication environment with real-time dynamic conditions.
Thus, it is essential to find new efficient mathematical solution methods to solve the tricky issues. 
With the powerful learning capability of deep reinforcement learning (DRL), the adjustable parameters can be dynamically changed in real time according to the change of environmental conditions, ensuring the real-time results of the resource allocation scheme. 
Hence, DRL is introduced as an efficient intelligent resource allocation method. In recent years, 
DRL has been widely used to solve the large-scale non-convex resource allocation problems\cite{9372298,8714026,9500325}.

\vspace{-0.3cm}
\subsection{Related Work}
\subsubsection{\textit{Resource Allocation for CRN}}
CR can exploit spectrum idle opportunity for dynamic spectrum access, thus greatly improving spectrum efficiency. 
Resource allocation problems are considered in three main paradigms of CR networks (CRNs), including spectrum sharing, opportunistic spectrum access and sensing-enhanced spectrum sharing.
The transmit power of the secondary base station (SBS) was adjusted based on received signal strengths at the sensors under the spectrum sharing mechanism in \cite{9046301}.
In \cite{8115264}, multiband cooperative spectrum sensing was proposed, and an optimization problem was presented to minimize the energy consumption.
The authors proposed cross-layer reconfiguration schemes to address the resource allocation problem under the opportunistic spectrum access, considering the quality-of-service (QoS) requirement.
Under the opportunistic spectrum access mechanism and sensing-enhanced spectrum sharing mechanism, the authors in \cite{7946279} formulated a resource allocation problem with power transfer, which aimed to achieve robust max$\raisebox{0mm}{-}$min fairness.
Considering the worst-case channel state information (CSI) and imperfect spectrum sensing, the assignment of subchannels, the sensing time and the transmit power were jointly optimized to maximize the sum throughput of the secondary network.
The simulation indicated that the secondary network under the mechanism of sensing-enhanced spectrum sharing outperformed that under the mechanism of opportunistic spectrum access, along with the intricacy of the implementation.
Thus, CRNs with sensing-enhanced spectrum sharing are introduced for high spectral efficiency in this paper.
Moreover, more spectrum-efficient and energy-efficient approaches need to be further explored for 6G communications.

\subsubsection{\textit{Resource Allocation for IRS-assisted CRN}}
IRS has attracted widespread attention due to its ability to enchace the spectral efficiency in an energy-efficient method.
In \cite{wu2023joint}, the authors proposed an IRS-enhanced CRN under the opportunistic spectrum access mechanism. 
A block coordinate descent (BCD) based-algorithm was proposed to maximize the achievable rate of the secondary network.
An IRS-assisted wideband CRN with the mechanism of sensing-enhanced spectrum sharing was introduced in \cite{wu2021resource}.
To achieve the maximum throughput of the secondary network, the assignment of subchannels, the sensing time, the transmit power of the SBS and the reflection coefficients of IRS were jointly optimized.
The authors in \cite{9183907} formulated a promblem of maximizing spectral efficiency in an IRS-assisted full-duplex CRN, and a BCD-based resource allocation scheme was proposed.
However, most of the above works only focused on resource allocation problems with the absence of Eves and did not consider secure transmission.

\subsubsection{\textit{Resource Allocation for IRS-assisted PLS}}
Besides the enhancement of spectral efficiency, IRS has been widely used in secure transmission owing to its ability to suppress eavesdropping and enhance the received signal of legitimate users by the strengthened beamforming.
In \cite{9279316,9428001,9807309}, the secrecy rate was improved with implements of the IRS. 
In \cite{9279316}, an IRS-enhanced PLS of downlink wireless communication was considered, and the authors designed an algorithm with fractional programming to determine the optimal transmit beamforming under the fixed IRS phase shifts.
However, the works \cite{9279316,9428001,9807309} only employed IRS to enhance the secure transmission and failed to further exploit the spectrum efficiency.
In \cite{9389801} and \cite{9520295}, IRS-assisted secure communication was explored in multi-user multiple-input multiple-output (MIMO) systems with OFDM.
The reflection coefficients of IRS and transmit power of the BS were jointly designed in \cite{9389801}, and a BCD method was applied to address the resource allocation problems for secure communication.
To further improve spectrum utilization and broaden application scenarios, CRNs were considered in the IRS-assisted secure transmission schemes \cite{9542962,9148890}. 
For the purpose of maximizing secrecy energy efficiency, the authors in \cite{9542962} jointly designed the coefficients of IRS reflection elements and transmit beamforming, and proposed an iterative penalty function based algorithm.
Note that majority of the aforementioned studies used traditional mathematical approaches to address the complex non-convex resource allocation issues, which is difficult to fulfill the requirements of real-time performance, particularly for communication systems of large scale.

\subsubsection{\textit{Resource Allocation Based on DRL}}
Due to the fact that DRL can solve the complex non-convex problems quickly and achieve outstanding real-time performance, it has been widely used as an intelligent optimization approach for resource allocation \cite{wu2022intelligent,9448143,9759480,9206080}.
A hybrid DRL based resource allocation scheme was investigated in \cite{wu2022intelligent} for IRS-enhanced communication systems with OFDM.
However, this work considered only a single IRS and did not take into account the sensing-enhanced spectrum sharing and secure communication.
The authors in \cite{9448143} studied the decentralized contention-based spectrum sharing and proposed a distributed reinforcement learning (RL) scheme to maximize the downlink throughput. 
In \cite{9759480}, the authors considered a OFDM-based CRN with dynamic spectrum access. The successful access rate and collision rate were improved by the deep recurrent Q-network (DRQN) algorithm.
Moreover, in \cite{9206080}, the beamforming of BS and the reflection coefficients of IRS were jointly optimized to enhance secure transmission in IRS-enhanced communication networks.
An intelligent algorithm utilizing the post-decision state was exploited for the optimal resource allocation policy.
In \cite{9508416}, the authors maximized the throughput and secrecy rate under the delay constraints. 
A multi-agent DRL scheme was presented to tackle the challenge problems.
However, most of the above studies employed only one IRS, which could not achieve wide coverage and great performance improvement for all users in large-scale communication systems.
In \cite{9473585}, a multi-IRS-assisted MIMO communication network under imperfect CSI was introduced. 
To achieve the maximum sum rate, the IRS reflection elements, transmit power and BS combiners were jointly optimized.
DRL-based schemes were proposed to address the challenging resource allocation problem. However, secure communication has not been considered in this work.

\vspace{-0.2cm}
\subsection{Motivation and Contributions}
Most of the afore-mentioned works proposed resource allocation schemes for improving spectral efficiency and secure performance by using the traditional mathematical methods, 
which are usually accompanied by complex computational processes, making it difficult to obtain the real-time performance. 
Moreover, the existing studies applying intelligent schemes based on DRL algorithms have only focused on spectrum sharing for a single IRS or transmission enhancement for multiple IRSs, 
but have not focused on joint optimization of secure transmission and efficient spectrum utilization in multi-IRS-assisted spectrum sharing networks.

In this paper, secure communication is considered in the multi-IRS-assisted sensing-enhanced wideband spectrum sharing network. 
An intelligent resource allocation scheme by exploiting the hybrid hierarchical DRL is proposed to tackle the tricky non-convex resource allocation problem with coupled variables and integer programming. 
The primary contributions of this paper can be summarized as follows.
\begin{itemize}
  \item It is the first time that secure transmission is studied in the multi-IRS-assisted sensing-enhanced wideband spectrum sharing network. 
  The secrecy rate of the secondary network is maximized by jointly optimizing the transmit beamforming of the SBS, IRS pairing, subchannel assignment, reflection coefficients of IRSs and the sensing time.
  Multiple IRSs are utilized for both the sensing phase and transmission phase to optimize the detection probability and enhance secure transmission.
  \item An intelligent resource allocation scheme by exploiting the hybrid hierarchical DRL for secure transmission is designed to address the tricky non-convex problem with coupled variables and integer programming. 
  To tackle the dimensional disaster problem caused by massive reflective elements of multiple IRSs and the mixed action space problem, a hybrid hierarchical D3QN-soft Actor-Critic (SAC) framework is proposed.
  In particular, D3QN is exploited to address discrete action allocation while SAC is used to address continuous action space, and IRS pairings are abstracted as an option to alleviate the sparse reward problem. 
  \item An alternative optimization (AO)-based traditional mathematical method is considered to highlight the computational efficiency of our proposed intelligent algorithm.
  Simulation results demonstrate that our proposed scheme outperforms benchmark schemes in terms of the achievable secrecy rate of the secondary network.
  Moreover, our proposed hybrid hierarchical D3QN-SAC (H2DS) framework can efficiently tackle the sparse reward problem and dimensional disaster caused by large numbers of reflection elements of multiple IRSs.
  It is also shown that the inappropriate deployment of IRSs can reduce the security performance when Eves are present. 
\end{itemize}

\vspace{-0.4cm}
\subsection{Organization and Notation}
The subsequent sections of the paper is organized as follows. 
Section II introduces a multi-IRS-assisted sensing-enhanced wideband spectrum sharing network with multiple Eves. 
Section III presents the resource allocation problem.
Section IV proposes an intelligent resource allocation scheme for jointly optimizing the IRS pairing, subchannel assignment, reflection coefficients, transmit beamforming and sensing time.
Section V presents the simulation results and the analysis of the results. 
Finally, Section VI draws conclusions.

In this paper, italic letters denote scalars while bold-face letters denote vectors and matrices. 
For a vector $\mathbf{v}$, $\operatorname{diag}(\mathbf{v})$ denotes a diagonal matrix and $\mathbf{v}^T$ denotes the transposition of the $\mathbf{v}$. 
Let $\{\cdot\}$ represent a set of elements, $\mathbb{C}^{x \times y}$ denote the space of $x \times y$ complex-valued matrices, $\mathbb{E}[\cdot]$ represent the mathematic expectation, and $\arg \underset{a}{{\max}}f(a)$ denote the value of $a$ when $f(a)$ takes its maximum value.

\section{System Model}
\begin{figure}
\centering
\includegraphics[scale=0.55]{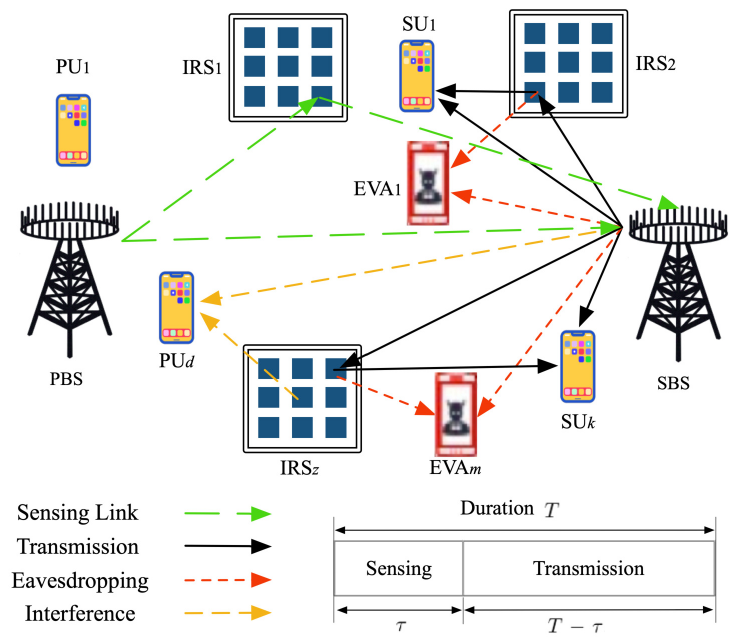}
\caption{A multi-IRS-assisted sensing-enhanced wideband secure communication network.}
\vspace{-0.5cm}
\end{figure}

A multi-IRS-assisted sensing-enhanced wideband spectrum sharing network with the presence of eavesdropping is considered as illustrated in Fig. 1, which includes a primary network, a secondary network and an eavesdropping network.
The primary network consists of a primary base station (PBS) with $N_p$ antennas and $D$ primary users (PUs), each with a single antenna.
The secondary network comprises a SBS with $N_s$ antennas and $K$ secondary users (SUs), each with a single antenna.
The spectrum of the primary network can be shared by the secondary network.
The PU set is denoted by $\mathcal{D}=\{1, \cdots, D\}$, and SU set is represented by $\mathcal{K}=\{1, \cdots, K\}$.
The eavesdropping network comprises $M$ Eves, each with a single antenna and a task of eavesdropping on SUs. 
The Eve set is written as $\mathcal{M}=\{1, \cdots, M\}$.
The wideband spectrum is divided into multiple narrow spectrum bands, 
and the subchannel set is represented as $\mathcal{C}=\{1, \cdots, C\}$. 

Multiple IRSs are around the SBS, and each IRS is equipped with $N_n$ passive reflection elements.
The reflection coefficients of IRSs can be adaptively adjusted by the IRS central control unit to optimize the network performance. 
Let $\mathcal{Z}=\{1, \cdots, Z\}$ denote the IRS set, and $\mathbf{\Phi}_z=\operatorname{diag}\left(b_{z,1} e^{j \phi_{z,1}}, b_{z,2} e^{j \phi_{z,2}}, \cdots, b_{z,N_n} e^{j \phi_{z,N_n}}\right) \in \mathbb{C}^{N_{n} \times N_{n}}$ represent the phase shift
matrix of the $z$-th IRS, in which $b_{z,n} \in$ $[0,1]$ denotes the $n$-th amplitude correlation coefficient of the $z$-th IRS, and $\phi_{z,n} \in[0,2 \pi)$ denotes the phase correlation coefficient of the $z$-th IRS. 

To enhance the spectral efficiency, the availability of each subchannel in the primary network is detected by energy detection-based spectrum sensing that is employed at the SBS.
The SBS is capable of dynamically adjusting the transmit power based on the spectrum sensing results, while simultaneously maintaining the required QoS for the primary network.
The temporal structure of the communication frame is denoted by $T$, which includes two phases: the sensing phase with a duration of $\tau$ and the data transmission phase with a duration of $T-\tau$.

Let $\mathbf{H}_{z} \in \mathbb{C}^{N_n \times N_p}$, $\mathbf{g}_{z, s} \in \mathbb{C}^{N_n \times N_s}$ and $\mathbf{h}_{p, s} \in \mathbb{C}^{N_p \times N_s}$ respectively denote the channels from the PBS to the $z$-th IRS, from the $z$-th IRS to the SBS and from the PBS to the SBS.
Hence, according to \cite{choi2013optimal}, the probability of detection $P_{d}$ and false alarm $P_{f}$ obtained based on energy detection can be respectively given by
\begin{subequations}\label{r1}
\begin{align}
&P_{d}(c,z,\boldsymbol{\Phi}_z,\tau)=Q\left(\frac{\epsilon-\sigma_s^2(N_s+\gamma_{z,c})}{\left(\sigma_s^2 / \sqrt{\tau f_s}\right) \sqrt{2 \gamma_{z,c}+N_s}}\right) \tag{\ref{r1}a},\\
&P_{f}(c,z,\boldsymbol{\Phi}_z,\tau)=Q\left(\frac{\epsilon-\sigma_s^2 N_s}{\left(\sigma_s^2 / \sqrt{\tau f_s}\right) \sqrt{N_s}}\right) \tag{\ref{r1}b},
\end{align}
\end{subequations}
where $f_s$ is the sampling frequency.
Let $\mathbf{f}_{c}^p$ denote the transmit beamforming of the PBS over the subchannel $c$,
and $\gamma_{z,c}$ denote the sensing SNR at the SBS on the $c$-th subchannel assisted by the $z$-th IRS,
which can be expressed as
\begin{equation}
\gamma_{z,c}=\left\|\mathbf{h}_{\mathrm{p,s}}^H+\mathbf{g}_{z,s}^H \boldsymbol{\Phi}_z \mathbf{H}_{z}\right\|^2 \alpha_{c},
\end{equation}
where $\alpha_{c}=\left\| \mathbf{f}_{c}^p\right\|^2 / \sigma_s^2$ when the $c$-th subchannel is occupied by the PBS; otherwise, $\alpha_{c}=0$.
The function $Q(\cdot)$ represents the the complement of the cumulative distribution function for a standard Gaussian random variable,
expressed as
\begin{equation}
Q(x)=\frac{1}{\sqrt{2 \pi}} \int_x^{\infty} \exp \left(-\frac{t^2}{2}\right) d t.
\end{equation}

The selection of the detection threshold $\epsilon$ is determined to achieve a specific probability of detection, as given in equation (\ref{r1}a), which can be expressed as
\begin{equation}\label{eq4}
\bar{\epsilon}=\left(Q^{-1}\left(\bar{P}_{d}\right) \sqrt{\frac{2 \gamma_{z,c}+N_s}{\tau f_s}}+\gamma_{z,c}+N_s\right) \sigma_s^2,
\end{equation}
where $Q^{-1}(\cdot)$ refers to the inverse of the $Q$-function, and $\bar{P}_{d}$, a given value, is the target detection probability. By utilizing the threshold in (\ref{eq4}), the false alarm probability (\ref{r1}b) is rewritten as
\begin{equation}
\bar{P}_{f}(c,z,\boldsymbol{\Phi}_z,\tau)=Q\left(\sqrt{\frac{2 \gamma_{z,c}+1}{N_s}} Q^{-1}\left(\bar{P}_{d}\right)+\sqrt{\frac{\tau f_s}{N_s}} \gamma_{z,c}\right) .
\end{equation}

Let $\operatorname{Pr}\left(\mathcal{H}_c^0\right)$ and $\operatorname{Pr}\left(\mathcal{H}_c^1\right)$ respectively denote the probabilities of the $c$-th subchannel being available ($\mathcal{H}_c^0$) and being occupied ($\mathcal{H}_c^1$). 
Let $\operatorname{P}(\cdot)$ represent all probability of the real state and the sensing state of the $c$-th subchannel with the $z$-th IRS, which can be respectively expressed as
\begin{subequations}\label{reg2}
\begin{align}
  &\operatorname{P}^{00}(c,z,\boldsymbol{\Phi}_z,\tau)=\operatorname{Pr}\left(\mathcal{H}_c^0\right)\left(1-\bar{P}_{f}(c,z,\boldsymbol{\Phi}_z,\tau)\right), \tag{\ref{reg2}a}\\
  &\operatorname{P}^{01}(c,z,\boldsymbol{\Phi}_z,\tau)=\operatorname{Pr}\left(\mathcal{H}_c^1\right)\left(1-\bar{P}_{d}(c,z,\boldsymbol{\Phi}_z,\tau)\right), \tag{\ref{reg2}b} \\
  &\operatorname{P}^{10}(c,z,\boldsymbol{\Phi}_z,\tau)=\operatorname{Pr}\left(\mathcal{H}_c^0\right) \bar{P}_{f}(c,z,\boldsymbol{\Phi}_z,\tau), \tag{\ref{reg2}c} \\
  &\operatorname{P}^{11}(c,z,\boldsymbol{\Phi}_z,\tau)=\operatorname{Pr}\left(\mathcal{H}_c^1\right) \bar{P}_{d}(c,z,\boldsymbol{\Phi}_z,\tau), \tag{\ref{reg2}d}
\end{align}
\end{subequations}
where the double-digit superscript indicates the index of the spectrum sensing result provided by the SBS and the index of the actual state of the subchannel, respectively. 
Let “0” represent an idle subchannel and “1” represent an occupied subchannel.

Let $\mathbf{g}_{z,d} \in \mathbb{C}^{N_n \times 1}$ and $\mathbf{h}_{p,d} \in \mathbb{C}^{N_p \times 1}$ respectively represent the channels from the $z$-th IRS to the $d$-th PU and from the PBS to the $d$-th PU. 
Let $\mathbf{G}_{z} \in \mathbb{C}^{N_n \times N_s}$ and $\mathbf{h}_{s,k} \in \mathbb{C}^{N_s \times 1}$ respectively represent the channels from the SBS to the $z$-th IRS and from the SBS to the $k$-th PU. 
The spectrum band of each subchannel is set to $1\mathrm{~MHz}$.
Let $\mathbf{f}_{d,c}^p$ denote the beamforming vector from the PBS to the $d$-th PU over the $c$-th subchannel and the PBS transmits the same power over each subchannel. 
Let $\mathbf{f}_{k,c}^s$ denote the beamforming vector from the SBS to the $k$-th PU over the $c$-th channel. 
Considering the subchannel occupancy and QoS requirement of the PUs, the SBS can dynamically adjust the transmit power over each subchannel. 

In the transmission stage, two typical cases must be taken into account. 
Case one is that the $c$-th subchannel is practically inactive ($\mathcal{H}_c^0$). 
In this case, the received signal at the $k$-th SU over the $c$-th subchannel is written as
\begin{equation}
    y_{k,z,c}^0=\left(\mathbf{h}_{\mathrm{s}, \mathrm{k}}^T+\mathbf{g}_{z, k}^T \boldsymbol{\Phi}_{z} \mathbf{G}_{z}\right) \mathbf{f}_{k,c}^s s_{k}+n_{k},
\end{equation}
where $s_{k} \sim \mathcal{C N}(0,1)$ represents the data symbol transmitted from the SBS to the $k$-th SU, and $n_{k} \sim \mathcal{C N}\left(0, \sigma_{k}^2\right)$ is the additive white Gaussian noise (AWGN) at the $k$-th SU.
The received signal at the $m$-th Eve eavesdropping on the $k$-th SU over the $c$-th subchannel can be expressed as
\begin{equation}
    y_{m,k,c}^0=\left(\mathbf{h}_{s, m}^T+\mathbf{g}_{z_k, m}^T \boldsymbol{\Phi}_{z_k} \mathbf{G}_{z_k}\right)  \mathbf{f}_{k,c}^{s} s_{k} + n_{m},
\end{equation}
where $n_{m} \sim \mathcal{C N}\left(0, \sigma_{m}^2\right)$ is the AWGN at the $m$-th Eve.
The subscript $z_k$ represents the IRS that is paired with the $k$-th SU for secure transmission.

In case two, the $c$-th subchannel is occupied ($\mathcal{H}_c^1$), while there is a miss detection. 
Hence, the received signal at the $k$-th SU is expressed as
\begin{equation}
  \begin{aligned}
    y_{k,z,c}^1=&\left(\mathbf{h}_{\mathrm{s}, \mathrm{k}}^T+\mathbf{g}_{z, k}^T \boldsymbol{\Phi}_{z} \mathbf{G}_{z}\right) \mathbf{f}_{k,c}^s s_{k} \\
    &+\sum_{d=1}^D \delta_{d,c}\left(\mathbf{h}_{p, k}^T+\mathbf{g}_{{z_d}, k}^T \boldsymbol{\Phi}_{z_d} \mathbf{H}_{z_d}\right) \mathbf{f}_{d,c}^p s_{d}+n_{k},
    \end{aligned}
\end{equation}
where $\delta_{d,c}$ indicates whether the $d$-th PU uses the $c$-th subchannel for transmission or not. If the $d$-th PU occupies the $c$-th subchannel, $\delta_{d,c}=1$; otherwise, $\delta_{d,c}=0$. Then, the received signal at the $m$-th Eve which eavesdrops on the $k$-th SU over the $c$-th subchannel can be written as
\begin{equation}
  \begin{aligned}
    y_{m,k,c}^1=&\left(\mathbf{h}_{s, m}^T+\mathbf{g}_{{z_k}, m}^T \mathbf{\Phi}_{z_k} \mathbf{G}_{z_k}\right)  \mathbf{f}_{k,c}^s s_{k} \\
    &+\sum_{d=1}^D \delta_{d,c}\left(\mathbf{h}_{p, m}^T+\mathbf{g}_{{z_d}, m}^T \boldsymbol{\Phi}_{z_d} \mathbf{H}_{z_d}\right) \mathbf{f}_{d,c}^p s_{d}+n_{m},
  \end{aligned}
\end{equation}
where the subscript $z_d$ and $c_d$ represent the IRS and the channel used for the transmission from the PBS to the $d$-th PU, respectively.

Furthermore, noting the Qos of PUs, the received signal at the $d$-th PU over the $c$-th subchannel can be represented as
\begin{equation}
  \begin{aligned}
  y_{d,z,c}=&\left(\mathbf{h}_{p, d}^T+\mathbf{g}_{z, d}^T \boldsymbol{\Phi}_{z} \mathbf{H}_{z}\right) \mathbf{f}_{d,c}^p s_{d}\\
  &+\sum_{k=1}^K \xi_{k,c}\left(\mathbf{h}_{s,d}^T+\mathbf{g}_{{z_k}, d}^T \boldsymbol{\Phi}_{z_k} \mathbf{G}_{z_k}\right) \mathbf{f}_{k,c}^s s_{k}+n_{d},
  \end{aligned}
\end{equation}
where $\xi_{k,c}$ indicates the channel allocation state. If the $k$-th SU occupies the $c$-th subchannel, $\xi_{k,c}=1$; otherwise, $\xi_{k,c}=0$.

As a result, the instantaneous transmission rate of the $k$-th SU paired with $z$-th IRS over the $c$-th subchannel can be respectively given by 
\begin{subequations}\label{reg7}
  \begin{align}
    &r_{k}^{00}\left(c,z,\boldsymbol{\Phi}_z,\mathbf{f}^s_{k,c},\tau\right)= \left(1-\frac{\tau}{T}\right) \log _2\left(1+\operatorname{SINR}_{k}^{00}\right),\\
    &r_{k}^{10}\left(c,z,\boldsymbol{\Phi}_z,\mathbf{f}^s_{k,c},\tau\right)= \left(1-\frac{\tau}{T}\right) \log _2\left(1+\operatorname{SINR}_{k}^{10}\right),\\
    &r_{k}^{01}\left(c,z,\boldsymbol{\Phi}_z,\mathbf{f}^s_{k,c},\tau\right)= \left(1-\frac{\tau}{T}\right) \log _2\left(1+\operatorname{SINR}_{k}^{01}\right),\\
    &r_{k}^{11}\left(c,z,\boldsymbol{\Phi}_z,\mathbf{f}^s_{k,c},\tau\right)= \left(1-\frac{\tau}{T}\right) \log _2\left(1+\operatorname{SINR}_{k}^{11}\right),
  \end{align}
\end{subequations}
where the double-digit superscript denotes the index of the spectrum sensing result made by the SBS and the index of the actual state of the subchannel, respectively. 
Let ${\rm SINR}_{k}$ represent the signal-to-interference-and-noise ratio (SINR) at the $k$-th SU over the $c$-th subchannel using the $z$-th IRS,
which are respectively written as
\begin{subequations}
  \begin{align}
    &{\rm SINR}_{k}^{00}=\frac{\left|\left(\mathbf{h}_{s, k}^T+\mathbf{g}_{{z}, k}^T \mathbf{\Phi}_{z} \mathbf{G}_{s, R_{z}}\right) \mathbf{f}_{k,c}^s\right|^2}{\sigma_{k}^2},\\
    &{\rm SINR}_{k}^{10}=\frac{\left|\left(\mathbf{h}_{s, k}^T+\mathbf{g}_{{z}, k}^T \mathbf{\Phi}_{z} \mathbf{G}_{s, R_{z}}\right) \mathbf{f}_{k,c}^s\right|^2}{\sigma_{k}^2},\\
    &{\rm SINR}_{k}^{01}=\frac{\left|\left(\mathbf{h}_{s, k}^T+\mathbf{g}_{{z}, k}^T \mathbf{\Phi}_{z} \mathbf{G}_{s, R_{z}}\right) \mathbf{f}_{k,c}^s\right|^2}{\sum\limits_{d=1}^D \delta_{d,c}\left|\left(\mathbf{h}_{p, k}^T+\mathbf{g}_{{z_d}, k}^T \mathbf{\Phi}_{z_d} \mathbf{H}_{p, R_{z_d}}\right) {\mathbf{f}}_{d,c}^p\right|^2+\sigma_{k}^2},\\
    &{\rm SINR}_{k}^{11}=\frac{\left|\left(\mathbf{h}_{s, k}^T+\mathbf{g}_{{z}, k}^T \mathbf{\Phi}_{z} \mathbf{G}_{s, R_{z}}\right) \mathbf{f}_{k,c}^s\right|^2}{\sum\limits_{d=1}^D \delta_{d,c}\left|\left(\mathbf{h}_{p, k}^T+\mathbf{g}_{{z_d}, k}^T \mathbf{\Phi}_{z_d} \mathbf{H}_{p, R_{z_d}}\right) {\mathbf{f}}_{d,c}^p\right|^2+\sigma_{k}^2}.
  \end{align}
\end{subequations}

Once the $m$-th Eve attempts to eavesdrop on the signal of the $k$-th SU over the $c$-th subchannel, its instantaneous eavesdropping rate can be respectively expressed by
\begin{subequations}
  \begin{align}
    &r_{m,k}^{00}\left(c,z_k,\boldsymbol{\Phi}_{z_k},\mathbf{f}^s_{k,c},\tau\right)=\left(1-\frac{\tau}{T}\right) \nonumber \\
    &\log _2\left(1+\frac{\left|\left(\mathbf{h}_{s, m}^T+\mathbf{g}_{{z_k}, m}^T \mathbf{\Phi}_{z_k} \mathbf{G}_{z_k}\right) \mathbf{f}_{k,c}^s\right|^2}{\sigma_{m}^2}\right),\\
    &r_{m,k}^{10}\left(c,z_k,\boldsymbol{\Phi}_{z_k},\mathbf{f}^s_{k,c},\tau\right)=\left(1-\frac{\tau}{T}\right) \nonumber \\
    &\log _2\left(1+\frac{\left|\left(\mathbf{h}_{s, m}^T+\mathbf{g}_{{z_k}, m}^T \mathbf{\Phi}_{z_k} \mathbf{G}_{z_k}\right) \mathbf{f}_{k,c}^s\right|^2}{\sigma_{m}^2}\right),\nonumber \\
    &r_{m,k}^{01}\left(c,z_k,\boldsymbol{\Phi}_{z_k},\mathbf{f}^s_{k,c},\tau\right)= \left(1-\frac{\tau}{T}\right) \nonumber \\
    &\log _2\left(1+\frac{\left|\left(\mathbf{h}_{s, m}^T+\mathbf{g}_{{z_k}, m}^T \mathbf{\Phi}_{z_k} \mathbf{G}_{z_k}\right) \mathbf{f}_{k,c}^s\right|^2}{\sum\limits_{d=1}^D \delta_{d,c}\left|\left(\mathbf{h}_{p, m}^T+\mathbf{g}_{{z_d}, m}^T \mathbf{\Phi}_{z_d} \mathbf{H}_{z_d}\right) \mathbf{f}_{d,c}^p\right|^2+\sigma_{m}^2}\right),\\
    &r_{m,k}^{11}\left(c,z_k,\boldsymbol{\Phi}_{z_k},\mathbf{f}^s_{k,c},\tau\right)= \left(1-\frac{\tau}{T}\right) \nonumber \\
    &\log _2\left(1+\frac{\left|\left(\mathbf{h}_{s, m}^T+\mathbf{g}_{{z_k}, m}^T \mathbf{\Phi}_{z_k} \mathbf{G}_{z_k}\right) \mathbf{f}_{k,c}^s\right|^2}{\sum\limits_{d=1}^D \delta_{d,c}\left|\left(\mathbf{h}_{p, m}^T+\mathbf{g}_{{z_d}, m}^T \mathbf{\Phi}_{z_d} \mathbf{H}_{z_d}\right) \mathbf{f}_{d,c}^p\right|^2+\sigma_{m}^2}\right).
  \end{align}
\end{subequations}

Let $\zeta_{k,z}$ represent the pairing situation between the $z$-th IRS and the $k$-th SU. If the $k$-th user is paired with $z$-th IRS, $\zeta_{k,z}=1$; otherwise, $\zeta_{k,z}=0$. 
Let $\xi_{k}=\{\xi_{k,c}\}_{\forall c \in C}$ and $\zeta_{k}=\{\zeta_{k,z}\}_{\forall z \in Z}$ denote the channel allocation and IRS pairings of the $k$-th SU, respectively.
Thus, the target average rate of the $k$-th SU and the average eavesdropping rate of the $m$-th Eve eavesdropping the $k$-th SU can be respectively given as 
\begin{equation}
  \begin{aligned}
\bar{R}_{k}\left(\xi_{k},\zeta_{k}, \boldsymbol{\Phi}_z, \mathbf{f}^s_{k,c}, \tau\right) = &\sum_{c=1}^C \xi_{k,c}(\operatorname{P}^{00}r_{k}^{00}+\operatorname{P}^{10}r_{k}^{10}\\
&+\operatorname{P}^{01}r_{k}^{01}+\operatorname{P}^{11}r_{k}^{11}),
  \end{aligned}
\end{equation}
\begin{equation}
  \begin{aligned}
\bar{R}_{m,k}&\left(\xi_{k},\zeta_{k}, \boldsymbol{\Phi}_{z_k}, \mathbf{f}^s_{k,c}, \tau\right)=\sum_{c=1}^C \xi_{k,c}(\operatorname{P}^{00}r_{m,k}^{00}\\
&+\operatorname{P}^{10}r_{m,k}^{10}+\operatorname{P}^{01}r_{m,k}^{01}+\operatorname{P}^{11}r_{m,k}^{11}).
  \end{aligned}
\end{equation}

Let $\xi=\{\xi_{k}\}_{\forall k \in K}$, $\zeta=\{\zeta_{k}\}_{\forall k \in K}$, $\mathbf{\Theta}=\{\boldsymbol{\Phi}_z\}_{\forall z \in Z}$ and $\mathbf{F}=\{\mathbf{f}^s_{k,c}\}_{\forall k \in K,\forall c \in C}$
denote the channel allocation, the IRS pairings, the phase correlation coefficient of IRSs, the beamforming vector of SUs, respectively. 
The secrecy rate of the secondary network is expressed as
\begin{equation}
\bar{R}_s^{\sec}\left(\xi,\zeta,\mathbf{\Theta},\mathbf{F},\tau\right)=\sum_{k=1}^K\left(\bar{R}_{k}- \max_{\forall m} \bar{R}_{m,k}\right)^{+},
\end{equation}
where Eves can always reach the highest eavesdropping rate, and when the secrecy rate has negative values, the SBS stops information transmission and the secrecy rate keeps 0.

In view of the communication requirements of the PUs, the transmission rate of the $d$-th PU is represented by (18).
\begin{figure*}
  \begin{equation}
  \begin{aligned}
    \bar{R}_{d}(\xi,\zeta,\mathbf{\Theta},\mathbf{F},\tau)=P^{01}&\log _2\left[1+\frac{\left|\left(\mathbf{h}_{p, d}^T+\mathbf{g}_{{z^*}, d}^T \mathbf{\Phi}_{z^*} \mathbf{H}_{z^*}\right) \mathbf{f}_{d,c}^p\right|^2}{\sum\limits_{k=1}^K \xi_{k,c}\left|\left(\mathbf{h}_{s, d}^T+\mathbf{g}_{{z_k}, d}^T \boldsymbol{\Phi}_{z_k} \mathbf{G}_{z_k}\right) \mathbf{f}_{k,c}^s\right|^2+\sigma_{d}^2}\right] \\
    &+P^{11}\log _2\left[1+\frac{\left|\left(\mathbf{h}_{p, d}^T+\mathbf{g}_{{z^*}, d}^T \mathbf{\Phi}_{z^*} \mathbf{H}_{p, z^*}\right) \mathbf{f}_{d,c}^p\right|^2}{\sum\limits_{k=1}^K \xi_{k,c}\left|\left(\mathbf{h}_{s, d}^T+\mathbf{g}_{{z_k}, d}^T \boldsymbol{\Phi}_{z_k} \mathbf{G}_{z_k}\right) \mathbf{f}_{k,c}^s\right|^2+\sigma_{d}^2}\right].
  \end{aligned}
\end{equation}
\hrulefill
\end{figure*}
Since multiple IRSs can be used to assist the primary network,
$z^{*}$ denotes the $z$-th IRS that enables the PU to obtain the maximum transmission rate, and $\mathbf{\Phi}_{z^*}$ denotes the coefficients of the $z^*$-th IRS.

\section{Problem Formulation}
Based on the above discussions, the secrecy rate maximization problem in the multi-IRS-assisted sensing-enhanced wideband spectrum sharing network is formulated as
\begin{subequations}\label{pro}
  \begin{align}
  \mathbf{P1}: &\underset{\xi,\zeta,\mathbf{\Theta},\mathbf{F},\tau}{\operatorname{max}} \bar{R}_s^{\sec}\left(\xi,\zeta,\mathbf{\Theta},\mathbf{F},\tau\right) \\ 
  \text{ s.t. }  
  &\mathrm{C1}:\bar{R}_{d} \geq R_{d}^{\min}, \forall d\in \mathcal{D},\\
  &\mathrm{C2}:\bar{P}_{f} \leq \bar{P}_{f}^{\max},\\
  &\mathrm{C3}:\sum_{k=1}^K\sum_{c=1}^C\left\|\mathbf{f}_{k,c}^s\right\|^2 \leq TP_s,\\
  &\mathrm{C4}:\left|b_{z,n}e^{j \phi_{z,n}}\right|\leq 1, \forall z \in \mathcal{Z}, \forall n \in \mathcal{N},\\
  &\mathrm{C5}:b_{z,n} \in[0,1], \phi_{z,n} \in[0,2 \pi), \forall z \in \mathcal{Z}, \forall n \in \mathcal{N},\\
  &\mathrm{C6}:\xi_{k,c} \in\{0,1\},  \sum_{c=1}^{C} \xi_{k,c}=1,\sum_{k=1}^{K} \xi_{k,c}=1, \nonumber \\ 
  &\forall c \in\mathcal{C}, \forall k \in\mathcal{K},\\
  &\mathrm{C7}:\zeta_{k,z} \in\{0,1\}, \sum_{z=1}^{Z} \zeta_{k,z}=1, \forall z \in\mathcal{Z},\forall k \in \mathcal{K},
  \end{align}
\end{subequations}
where $\bar{P}_{f}^{\max}$ denotes the maximum tolerable probability of false alarm.
Constraint $\mathrm{C1}$ represents the requirement of the transmission rate of each PU.
Constraint $\mathrm{C2}$ denotes the constraint of the probability of false alarm.
$\mathrm{C3}$ means the constraint on the maximum transmission power $TP_s$ of the SBS.
$\mathrm{C4}$ and $\mathrm{C5}$ mean the constraints of the IRS reflection array elements.
Constraint $\mathrm{C6}$ ensures that each SU occupies a subchannel if the number of channels is sufficient and that there is no double occupancy between SUs.
Constraint $\mathrm{C7}$ ensures that there is only one match or no match between users and IRSs, and one user can only be paired with one IRS.
It can be seen that $\mathbf{P1}$ is a challenging non-convex problem with coupled variables and integer programming due to the joint optimization of discrete subchannel assignment, IRS paring, continuous IRS reflection coefficients and transmit beamforming.
Moreover, the large number of reflection elements of multiple IRSs causes the intractable dimension curse problem, making $\mathbf{P1}$ extremely difficult to tackle.

\vspace{-0.2cm}
\section{Our Proposed Intelligent Resource Allocation Scheme Using H2DS}
To tackle the challenging non-convex resource allocation problem with coupled variables and integer programming in eq. (\ref{pro}), 
an intelligent resource allocation scheme is designed by exploiting H2DS. 
The details are presented as follows. 

\vspace{-0.2cm}
\subsection{Problem Formulation Based on RL}
RL enables the agent to perceive and adapt to the dynamic environment and find the global optimal solutions, which can be described by Markov decision process (MDP). 
Here, our proposed resource allocation problem given by eq. (\ref{pro}) is formulated as a RL problem,
where the multi-IRS-assisted sensing-enhanced wideband spectrum sharing secure transmission network is considered as a dynamic environment, and the control center at SBS, which can access the sensors and IRS control units, is regarded as an intelligent agent. 
Besides the agent and the environment, the RL problem encompasses several other critical components that play a pivotal role in successful implementation, including the state space, action space, reward function, and transition probability.
Described below are the details of our designed components.

\textbf{State space:} The state space of the proposed multi-IRSs assisted sensing-enhanced wideband spectrum sharing network can be denoted by $\mathcal{S}$. The state $s_t \in \mathcal{S}$ includes the current channel information of SUs and Eves, 
previous selected actions, current achievable transmission rate of SUs, current achievable transmission rate of PUs, current eavesdropping rate of Eves and
the current achievable secrecy rate. Let $\mathbf{h}_{k, t}$ denote the channel coefficients of $k$-th SU at the current time $t$. 
Let $a_{t-1}$ denote the combination of selected discrete and continuous actions at the previous time $t-1$. At the current time $t$, the transmission rate of the $k$-th SU and the $d$-th PU is respectively denoted by $\bar{R}_{k, t}, \bar{R}_{d, t}$, the eavesdropping rate of the $m$-th Eve is denoted by $\bar{R}_{m, t}$, and the secrecy rate of the secondary network is denoted by $\bar{R}_{s, t}^{\mathrm{sec}}$. Hence, the state denote by $s_t$ at the timestamp $t$ can be written as
\begin{equation}
s_t=\left\{\left\{\mathbf{h}_{k, t}\right\}, a_{t-1},\left\{\bar{R}_{k, t}\right\},\left\{\bar{R}_{d, t}\right\},\left\{\bar{R}_{m, t}\right\}, \bar{R}_{s, t}^{\mathrm{sec}}\right\}.
\end{equation}

\textbf{Action space:} Let $\mathcal{A}$ denote the action space of the SBS. 
Here, the action $a^t \in A$ includes the assignments of subchannels, the pairings of IRSs, reflection coefficients of IRSs, transmit beamforming and the sensing time. 
Let $\xi_t$ and $\zeta_t$ represent the channel allocation and the IRS choice of the $k$-th SU at the current time $t$, respectively.
Let $\mathbf{\Theta}_t$, $\mathbf{F}_t$ and $\tau_t$ denote the reflection coefficients of the IRSs, the transmit beamforming of SUs, and the sensing time at the current time $t$, respectively.
Hence, the action $a_{t} \in \mathcal{A}$ can be defined by
\begin{equation}
a_t=\left\{\xi_t,\zeta_t,\mathbf{\Theta}_t,\mathbf{F}_t, \tau_t \right\}.
\end{equation}

\textbf{Transition probability:} Let $\mathcal{T}\left(s_{t+1} \mid s_t, a_t\right)$ denotes the probability of the network transitioning from a old state $s_t\in \mathcal{S}$ to a new state $s_{t+1}\in \mathcal{S}$ when the action $a_t \in \mathcal{A}$ is executed by the agent.

\textbf{Reward function:} The reward is a fundamental tool in the evaluation of intelligent algorithm to convey the training goal to the agent,
providing insight into the quality of the agent's action selection strategy.
Here, the reward function consists of the sum secrecy rate of the secondary network, secure transmission requirements of SUs and transmission rate requirements of PUs, which is designed as
\begin{equation}\label{cons}
r=\sum_{k=1}^K \left(\bar{R}_{k}- \max_{\forall m} \bar{R}_{m,k}\right)+ \nu_s \sum_{k=1}^K I_s+ \nu_d \sum_{d=1}^D I_d,
\end{equation}
where $\nu_s$ and $\nu_d$ are constant coefficients, $\bar{R}_{k}$ is the achievable rate of the $k$-th SU, and $\bar{R}_{m,k}$ is the rate of the $m$-th Eve eavesdropper the $k$-th SU. 
Let $I_k$ denote the punishment of the failure to achieve the minimum secure transmission rate of the $k$-th SU and $I_d$ denote the punishment of the failure to fulfill the data rate requirements of the $d$-th PU, which are respectively defined as
\begin{subequations}\label{cons1}
  \begin{align}
&I_s = \operatorname{min} \left\{\bar{R}_s^{\sec}-R_s^{\text {sec,min }},0\right\},\\
&I_d = \operatorname{min} \left\{\bar{R}_d-R_d^{\min},0\right\}.
  \end{align}
\end{subequations}
Using (\ref{cons1}), the intelligent agent can be penalized when failing to meet the minimum transmission requirements of PUs or failing to fulfill the minimum secrecy rate requirements for SUs, which ensures transmission requirements and alleviates the sparse reward problem caused by a large number of reflection coefficients of IRSs.

Let $\pi$ represent the policy, which can be regarded as action selection probabilities of the agent. Then, it has $\pi\left(s, a\right)$ displays the likelihood that the agent decides action $a$ in the state $s$.
The objective of the agent is to learn an optimal policy that maximizes the expected reward $r$ over a long term, which can be written as
\begin{equation}\label{reg10}
r(s,a)=\sum_{t=0}^{\infty} \gamma^t[r_{t+1}\mid (s_t,a_t)],
\end{equation}
in which $\gamma \in(0,1]$ denotes the discount factor, and $\gamma^t$ is intended for a more accurate estimate of the reward and varies exponentially with the iteration step $t$. 

In general, our designed optimal policy is expressed as
\begin{equation}
\pi^*\left(s, a\right)=\arg \underset{\pi}{{\max}} \, \mathbb{E}_\pi\left[ r\left(s, a\right)\right].
\end{equation}

\begin{figure*}
  \centering
  \subfigure[The framework for our proposed intelligent resource allocation scheme.]{
    \begin{minipage}{16cm}
    \includegraphics[width=\textwidth]{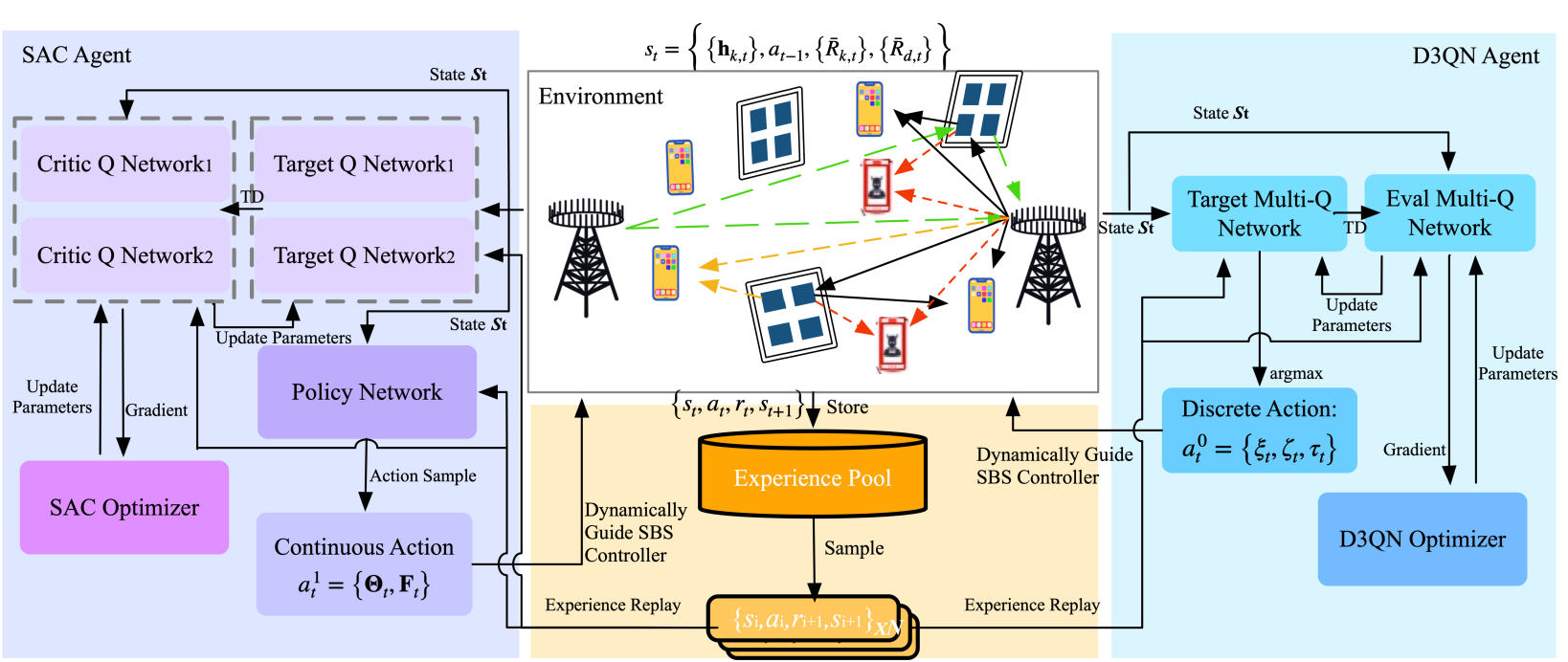} \\
    \end{minipage}}
  \subfigure[The details of networks exploited for our proposed resource allocation scheme.]{
    \begin{minipage}{16cm}
      \includegraphics[width=\textwidth]{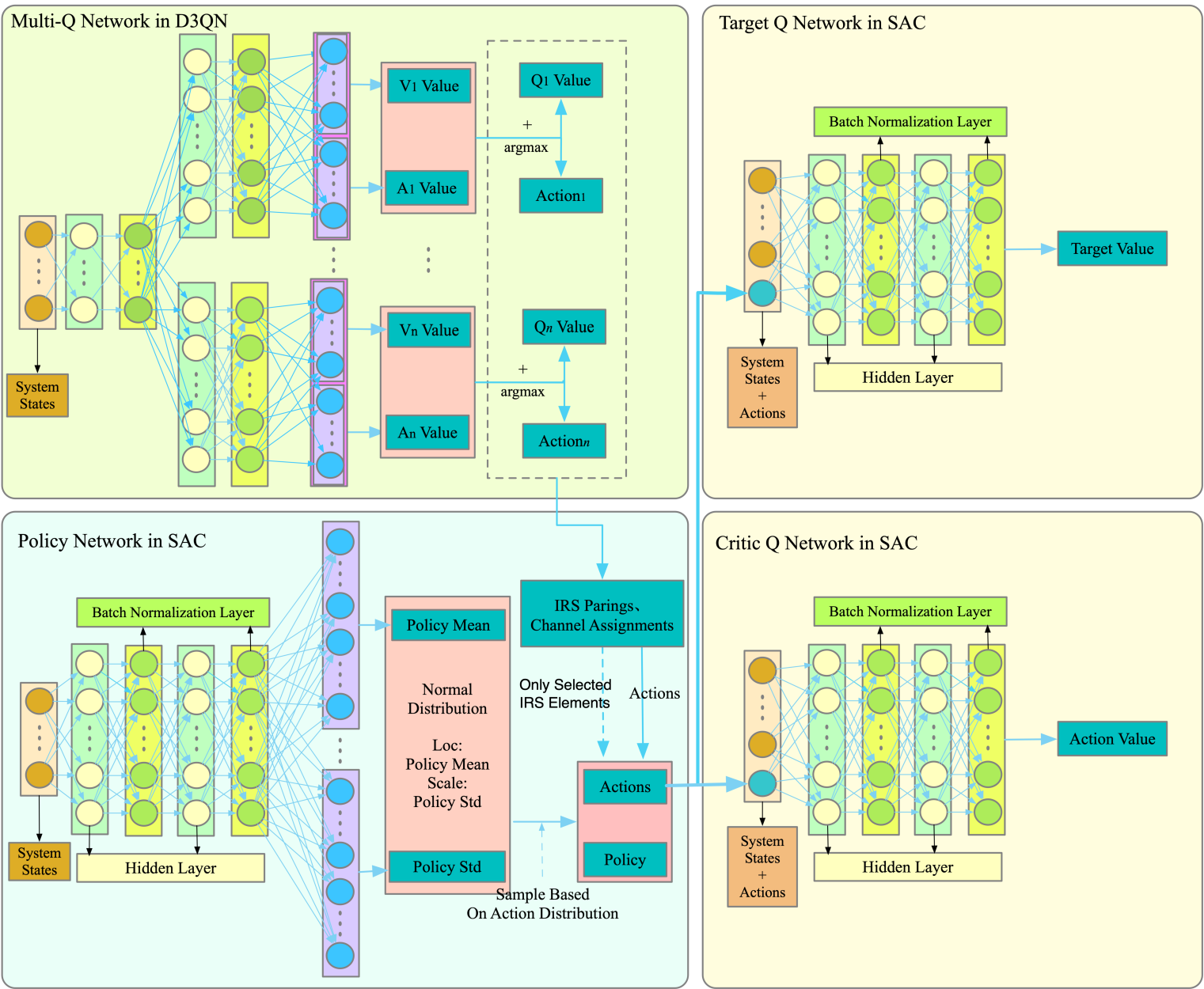} \\
    \end{minipage}}
  \caption{The proposed resource allocation scheme.}    
\end{figure*}

\vspace{-0.6cm}
\subsection{The Proposed Intelligent Resource Allocation Scheme Using D3QN-SAC}
Our proposed intelligent scheme for multi-IRS-assisted wideband spectrum sharing secure communication networks is given in Fig. 2.
The implementation details of our proposed intelligent scheme in communication networks are shown in Fig. 2(a), and the implementation details of deep networks in H2DS are shown in Fig. 2(b).
The framework of our proposed intelligent resource allocation scheme using D3QN-SAC is presented as follows.

\subsubsection{\textit{D3QN}} 
Combining the benefits of network design and architectural design in double deep Q-network (DQN) and dueling DQN, D3QN is able to efficiently handle high-bias problems and pay attention to the reward of the environmental state, which is suitable for determining the pairings of IRSs and assignments of subchannels.
D3QN uses the double Q network structure to eliminate the problem of overestimation and separates Q networks into two parts for separate calculation of state gains and action gains. 
One part of the Q network is only relevant to the state $s \in \mathcal{S}$, which is the value function $V$, while the other part of the Q network is solely relevant to the state $s \in \mathcal{S}$ and action $a \in \mathcal{A}$, which is the advantage function $A$.
The objective of D3QN is to maximize the Q value, which is represented as
\begin{equation}\label{q}
\begin{aligned}
Q_{\theta}(s, a) &=V_{\theta}(s) + \left(A_{\theta}(s, a)-\frac{1}{|\mathcal{A}|} \sum_{a^{\prime}} A_{\theta}\left(s, a^{\prime}\right)\right),
\end{aligned}
\end{equation}
where $\theta$ refers to the parameters of the eval Q network. 

\subsubsection{\textit{SAC}}SAC is a maximum entropy-based off-policy RL approach, which uses a random strategy to discover the best course of action and offers more chances to investigate the global optimal action than the deterministic method. 
In contrast to the conventional DRL learning objective, SAC not only considers the maximum long-term return but also needs to maximize the entropy of each output action generated by the policy, so as to achieve policy randomization. 
Moreover, SAC avoids slipping into suboptimal based on giving approximately equal probabilities to actions with comparable Q values and without giving a high probability to any option within the action range.
Due to the above mentioned advantages that SAC has, SAC is opted to decide for the high-dimensional continuous actions, solving the dimensional catastrophe and the sparse reward problem due to large numbers of reflective array elements.
The goal of SAC is to learn an optimal policy $\pi^*$, which is expressed as
\begin{equation}
\pi^*\left(s, a\right)=\arg \underset{\pi}{\operatorname{\max}} \, \mathbb{E}_\pi\left[\sum_t r\left(s, a\right)+\alpha \underbrace{H\left(\pi\left(\cdot \mid s\right)\right)}_{entropy}\right].
\end{equation}

\subsubsection{\textit{Hybrid Hierarchical DRL Based on D3QN-SAC}}
A hybrid hierarchical algorithm based on D3QN-SAC is presented for our proposed intelligent resource allocation scheme. 
To address the discrete actions and continuous actions in our proposed framework, a hybrid D3QN-SAC algorithm is proposed. 
Specifically, SAC is used to address the continuous actions, which include the reflection array elements of IRSs, the transmit beamforming and the sensing time. 
D3QN is used to deal with the discrete actions, which include the assignment of subchannels and pairings of IRSs.
Moreover, considering the dimensionality curse encountered by the RL problem, the hierarchical reinforcement learning is exploited for D3QN-SAC to tackle the high-dimensional continuous action issue caused by large numbers of reflective array elements of multiple IRSs 
by using the tense abstraction technique and action space decomposition technique.
Specifically, the assignment of subchannels and the pairing of IRSs are firstly abstracted as a set of cooperative actions. 
The agent selects a channel allocation set and an IRS pairing set as an option by using the D3QN. 
Given the IRS pairing set, only the reflection array elements of the paired IRS are required for the training phase of the SAC, which can effectively alleviate the dimensional disaster problem.

\vspace{-0.3cm}
\subsection{Architecture and Training Details}
\vspace{-0.1cm}
As illustrated in Fig. 2(b), the architecture of D3QN has one eval multi-Q network and one target multi-Q network,
and the architecture of SAC has one policy network, two critic Q networks and two target Q networks.
For each training round, $N$ samples are randomly sampled in the experience pool, in which one sample can be denoted as $\left(s_i, a_i, r_{i+1}, s_{i+1}\right)$.

The target value $y_i(\theta)$ in the D3QN networks is represented by
\begin{equation}\label{f}
  \begin{aligned}
  y_i(\theta)=r_{i}+\gamma Q^{\prime}_{\theta^{\prime}}&\left(s_{i+1}, \arg \underset{a^{i+1}}{\max} Q_{\theta}\left(s_{i+1}, a_{i+1}\right)\right),
  \end{aligned}
\end{equation}
where $\theta^{\prime}$ refers to the parameters of the target Q network.
Based on eq. (\ref{f}), the loss function of Q networks in D3QN is written as
\begin{equation}
L_Q\left(\theta\right) = \frac{1}{N} \sum_{i=1}^{N}\left(\mathrm{y}_{i}(\theta)-\mathrm{Q}_{\theta}\left(s_i, a_i\right)\right)^2.
\end{equation}

In the SAC networks, the target value is evaluated by the target networks, which can be represented as
\begin{equation}
y_i(w_{j})=r_i+\gamma \min _{j=1,2} Q_{\omega_j^{-}}\left(s_{i+1}, a_{i+1}\right)-\lambda \log \pi_\lambda\left(a_{i+1} \mid s_{i+1}\right),
\end{equation}
where $a_{i} \sim \pi_\lambda\left(\cdot \mid s_{i+1}\right)$. 
Hence, the loss function of the eval critic networks in the SAC can be written as
\begin{equation}\label{reg8}
L_Q(\omega_j)=\frac{1}{N} \sum_{i=1}^N\left(y_i(w_{j})-Q_{\omega_j}\left(s_i, a_i\right)\right)^2.
\end{equation}
Subsequently, once the action $\tilde{a}_i$ is obtained with the the trick of reparameterization, the policy network's loss function in the SAC is expressed as
\begin{equation}\label{reg9}
L_\pi(\lambda)=\frac{1}{N} \sum_{i=1}^N\left(\alpha \log \pi_\lambda\left(\tilde{a}_i \mid s_i\right)-\min _{j=1,2} Q_{\omega_j}\left(s_i, \tilde{a}_i\right)\right).
\end{equation}

Our proposed intelligent resource allocation scheme using the H2DS for the multi-IRS-assisted sensing-enhanced spectrum sharing secure transmission network is summarized in \textbf{Algorithm} \ref{alg:alg1}.
\begin{breakablealgorithm}\label{alg:alg1}
    \caption{Our Proposed Intelligent Resource Allocation Scheme}
    \begin{algorithmic}[1]
    \STATE Initialize system parameters: training episodes $N$, iteration steps $M$, training times $L$ and training batches $B$;
    \STATE Initialize eval network parameters: $\theta$ in D3QN, and $\omega_1$, $\omega_2$, $\lambda$ in SAC.
    \STATE Initialize target network parameters: $\theta \rightarrow \theta^{-}$, $\omega_1 \rightarrow \omega_1^{-}$, $\omega_2 \rightarrow \omega_2^{-}$, $\lambda \rightarrow \lambda^{-}$;
    \FOR{episode $n=1 \rightarrow N$}
        \STATE Random the intial action, $a_0=\left\{\xi_0, \zeta_0, \mathbf{\Theta}_0, \mathbf{F}_0, \tau_0\right\}$;
        \STATE Obtain the initial state with action $a_0$, $s_1=\biggl\{\left\{\mathbf{h}_{k,1}\right\}, a_{0}, \left\{\bar{R}_{d,1}\right\}, \left\{\bar{R}_{k,1}\right\}, \left\{\bar{R}_{m,1}\right\}, \bar{R}_{s,1}^{\mathrm{sec}}\biggr\}$;
            \FOR{step $m=1$ $\rightarrow$ $M$}
            \STATE Use $\epsilon-$ greedy policy in D3QN to decide the discrete action $\mathbf{a}_{t}^{1}$ in state $\mathbf{s}_{t}$;
            \STATE Use maximum entropy in SAC to decide the continuous action $a^{2}_t$ in state $\mathbf{s}_{t}$;
            \STATE Execute action $a_t=\left\{a^1_t, a^2_t\right\}$ and turn into next state $s_{t+1}$;
            \STATE Store the experience$\left(s_t, a_t, r_{t+1}, s_{t+1}\right)$ in the experience replay buffer;
            \FOR {time $l=1$ $\rightarrow$ $L$}
            \STATE Sample $B$ tuples $\left\{\left\{s_i, a_i, r_i, s_{i+1}\right\}\right\}_{i=1, \ldots, B}$ from the experience replay buffer;
            \STATE Computer the loss function respectively in SAC and D3QN;
            \STATE Generate gradient $\nabla_{\theta} Q_{\theta}$, $\nabla_{\omega_j} Q_{\omega_j}$ and $\nabla_{\lambda} \pi_\lambda$;
            \STATE Update eval network parameters with the gradient: $\theta$, ${\omega_j}$ and $\lambda$;
            \IF{$t$ \% $T$ == 0}
            \STATE Update target parameters, $\theta^{-}$ : $\theta \rightarrow \theta^{-}$, $\tau \omega_1+(1-\tau) \omega_1^{-}\rightarrow \omega_1^{-}$, $\tau \omega_2+(1-\tau) \omega_2^{-} \rightarrow \omega_2^{-}$;
            \STATE Update the coefficient of the entropy regularization term $\alpha$;
            \ENDIF
            \ENDFOR
        \ENDFOR
    \ENDFOR
    \end{algorithmic}
\end{breakablealgorithm}

\vspace{-0.2cm}
\section{Simulation Results}
This section presents an evaluation of the performance of our proposed intelligent resource allocation scheme for a multi-IRS-assisted sensing-enhanced spectrum sharing secure transmission network
using the H2DS, and compares it to benchmark schemes.

The simulation parameters of our scenario are set based on the parameters used in \cite{wu2022intelligent,wu2021resource} as follows, unless otherwise mentioned.
The number of antennas at the SBS is $M = 6$. 
The PBS and SBS are respectively located at $\left(300, 0, 50\right)$ and $\left(0, 0, 50\right)$. 
The maximum transmit power of the SBS is $TP_s = 30$ dBm.
There are $Z = 3$ IRSs which are deployed close to the SBS at $\left(0, 160, 0\right)$, $\left(150, 0, 0\right)$ and $\left(80, 80, 20\right)$, each with $N = 36$ reflection elements.
The number of the PUs is $D = 2$, and the PUs are located at $\left(270, 65, 0\right)$ and $\left(250, 10, 0\right)$.
The number of the SUs is $K = 2$, and the SUs are located at $\left(10, 150, 0\right)$ and $\left(130, 40, 0\right)$.
Similar to \cite{wu2022intelligent}, the channels from the SBS and PBS to the PUs and SUs undergo the Rayleigh fading while the channels from the SBS and PBS to the IRSs, as well as the channels from the IRSs to the PUs and SUs undergo the Rician fading,
which is designed to simulate the stochastic nature of the channel in the real environment. 
The number of the subchannels is set as $C = 2$.
The corresponding path fading can be expressed as $PL = \left(P L_0 - 10 \varsigma \log _{10}\left(d / d_0\right)\right)$ dB, where $PL_0 = 30$ dB, reference distance $d_0 = 1$ m and $\varsigma$ denotes the path loss exponent.
Specifically, the path loss exponents from the SBS and PBS to the users is set as $\varsigma_{bu}=3.75$, from the SBS and PBS to the IRSs is set as $\varsigma_{br}=2.2$, and
from the IRSs to the users is set as $\varsigma_{ru}=2.2$. 
The variance of the noise at the $k$-th SU is set as $\sigma_{k}^2 = 0.01$. 
The sampling frequency is $f_s =6$ MHz, while the target probability of detection and the maximum tolerable probability of the false alarm are $\bar{P}_d = 0.9$ and $\bar{P}_f = 0.1$, respectively. 
The inactive probability of the PBS over the $c$-th subchannel is set as $\operatorname{Pr}\left(\mathcal{H}_c^0\right) = 0.8$. 

In the proposed algorithm, 
the network architecture of D3QN has two evaluation networks and two target networks, each with three hidden layers, each layer having 128 neurons.
On the other hand, the network architecture of SAC has one policy network, two soft Q networks and two target Q networks, each with three hidden layers, each layer having 256 neurons. 
In the D3QN, the learning rate for the networks is $0.005$. 
In the SAC, the learning rate for the cross-entropy network is $0.004$, for the soft Q networks is $0.004$, and for the policy network is $0.004$. 
The size of the experience replay buffer that stores training experience is $U = 20000$. 

In addition to our proposed algorithm, benchmark schemes are presented for comparison:
\begin{itemize}
  \item \textbf{Without IRS}: The IRSs are no longer utilized for spectrum awareness and secure transmission.
  \item \textbf{Random Choice}: The pairing of the IRS is generated randomly, while the other parameters are generated using our proposed algorithm.
  \item \textbf{Fixed IRS}: The reflection coefficients of IRSs are fixed, while the other dynamic system parameters are generated using our proposed algorithm.
  \item \textbf{Opportunistic}: The SUs are only allowed to utilize the spectrum band of the PUs when the subchannel is sensed to be inactive.  
  \item \textbf{H2DT}: The D3QN algorithm and the twin delayed deep deterministic policy gradient (TD3) algorithm are the state-of-the-art (SOTA) DRL algorithms that addressed the problems formulated in in \cite{9838566} and \cite{NEURIPS2022}, respectively. Hence, the hybrid hierarchical D3QN-TD3 (H2DT) algorithm is introduced as the SOTA baseline in this paper for comparison.
  \item \textbf{H2DD}: The DQN algorithm and the deep deterministic policy gradient (DDPG) algorithm are the SOTA DRL algorithms that addressed the problems formulated in \cite{wu2022intelligent} and \cite{lyu2022efficient}, respectively. Thus, the hybrid hierarchical DQN-DDPG algorithm (H2DD) is considered as the SOTA baseline to compare to our proposed scheme.
  \item \textbf{AO Scheme}: In order to better compare the traditional mathematical methods with our proposed hybrid hierarchical intelligent algorithm, a mathematical scheme based on alternating optimization (AO) approach is proposed, further analyzing the two schemes in terms of computational complexity and secure performance through numerical simulation. The details of the processing can be seen in the \textbf{APPENDIX A}.
\end{itemize}

\vspace{-0.2cm}
\subsection{Algorithm Computational Complexity Comparison}
The absolute central processing unit (CPU) time per algorithm is used to reflect the computational complexity in a practical environment. 
Hence, numerical tests are performed on the laptop with a Four-core Intel Core i7 with the $2.30 \mathrm{~GHz}$ CPU and $16 \mathrm{~GB}$ DRAMs with LPDDR4X technology.  
The CPU times for performing those two schemes can be seen in \textbf{TABLE} I.
It is seen that our proposed intelligent scheme achieves 40-100 times more computational efficiency improvement in terms of CPU time compared to the AO-based traditional mathematical optimization scheme.
\begin{table}
	\centering
	\caption{CPU Time Comparison for Two Schemes}
\begin{tabular}{ccc}
  \hline 
  \hline 
  \multirow{2}*{Iteration Times} & \multicolumn{2}{|c}{CPU Time}\\
  \cline{2-3} 
  ~&\multicolumn{1}{|c}{D3QN-SAC Intelligent Scheme}&AO Scheme\\
  \hline 
  20& 11.43 ms& 446.43 ms\\
  \hline 
  40& 11.43 ms& 787.51 ms\\
  \hline 
  60& 11.43 ms& 1036.24 ms\\
  \hline
  \hline 
\end{tabular}
\vspace{-0.3cm}
\end{table}

\vspace{-0.2cm}
\subsection{Algorithm Reward and Convergence}
\begin{figure}
  \centering
  \includegraphics[scale=0.55]{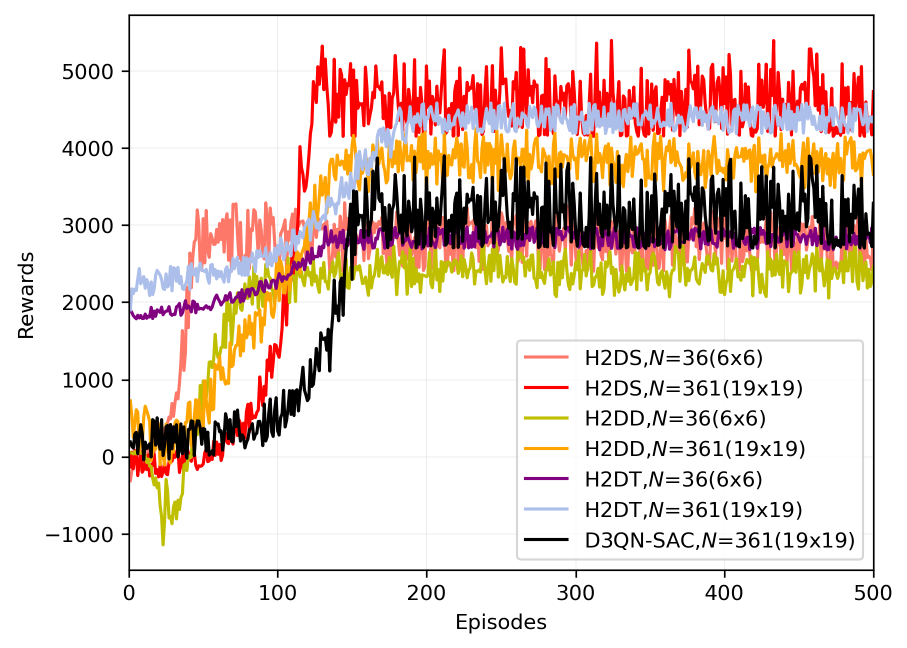}
  \vspace{-0.5cm}
  \caption{The convergence of different schemes.}
  \vspace{-0.4cm}
\end{figure}
Fig. 3 illustrates the convergence and achievable rewards of different intelligent algorithms versus episodes.
From Fig. 3, it can be observed that our proposed intelligent resource allocation scheme achieves convergence in short episodes and can efficiently capture the optimal strategy for resource allocation.
This is because our proposed H2DS algorithm can effectively alleviate the sparse reward problem caused by the dimensional catastrophe of massive reflection array elements, and successfully address the mixed action space problem.  
Furthermore, two baseline hybrid hierarchical algorithms are introduced for better comparison.
As depicted in Fig. 3, the H2DD algorithm exhibits a convergence rate approaching that of our proposed algorithm while failing to yield an outstanding reward.
This is due to the fact that the strong exploration ability plays a fundamental role in the case of multiple IRSs with high-dimensional continuous actions, and the H2DD is unable to address this issue.
Additionally, the H2DT algorithm shows good secure performance, but it faces challenges in reaching rapid convergence. 
The difficulty in convergence is attributed to the TD3 algorithm's susceptibility to training stability, especially in complex continuous action spaces and high-dimensional state spaces. 
Due to these challenges, the H2DT algorithm struggles to find the optimal direction efficiently for multi-IRS-assisted secure communication scenarios, resulting in slower convergence rates compared to other algorithms. 
Based on the experimental results and analysis, the HHDC algorithm is selected since it outperforms the baseline algorithms in terms of both convergence rate and rewards for our proposed communication network. 
Moreover, to explore the performance of algorithms with massive IRS reflective elements, the number of IRS elements is set to $N$=361. It is observed from the Fig. 3 that our proposed scheme outperforms other schemes even when multitudes of reflective elements are used.
Those results also demonstrate that our proposed hierarchical schemes perform better in addressing the high-dimension problem than the scheme without a hierarchical design.

\vspace{-0.3cm}
\subsection{Intelligent IRS Pairngs}
\begin{figure}
    \centering
    \setlength{\subfigcapskip}{-0.3cm}
    \subfigure[The proposed system model.]{
      \begin{minipage}{8cm}
      \includegraphics[width=\textwidth]{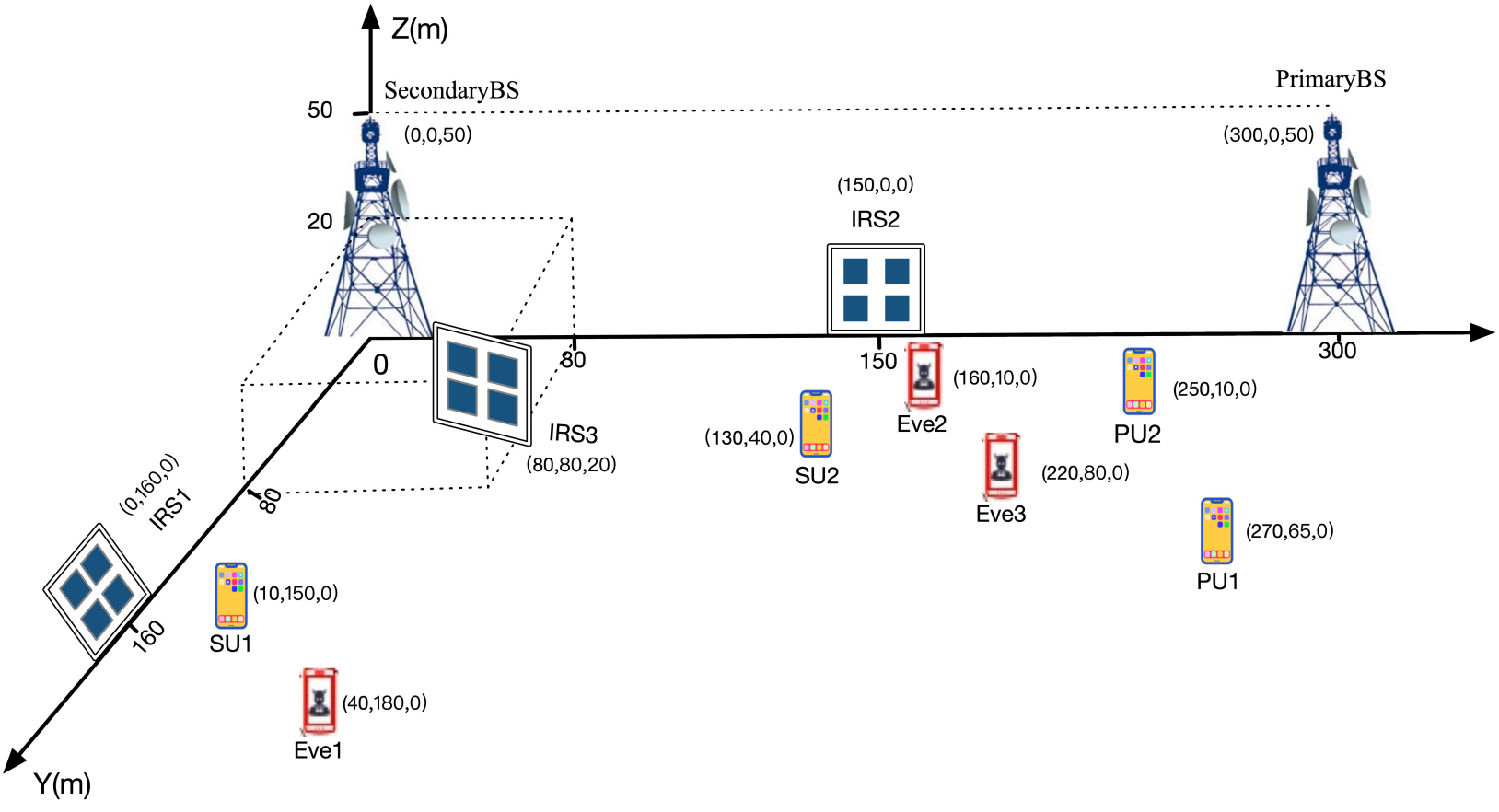} \\
      \end{minipage}}
    \subfigure[The table of IRS selection situation.]{
      \begin{minipage}{8cm}
        \includegraphics[width=\textwidth]{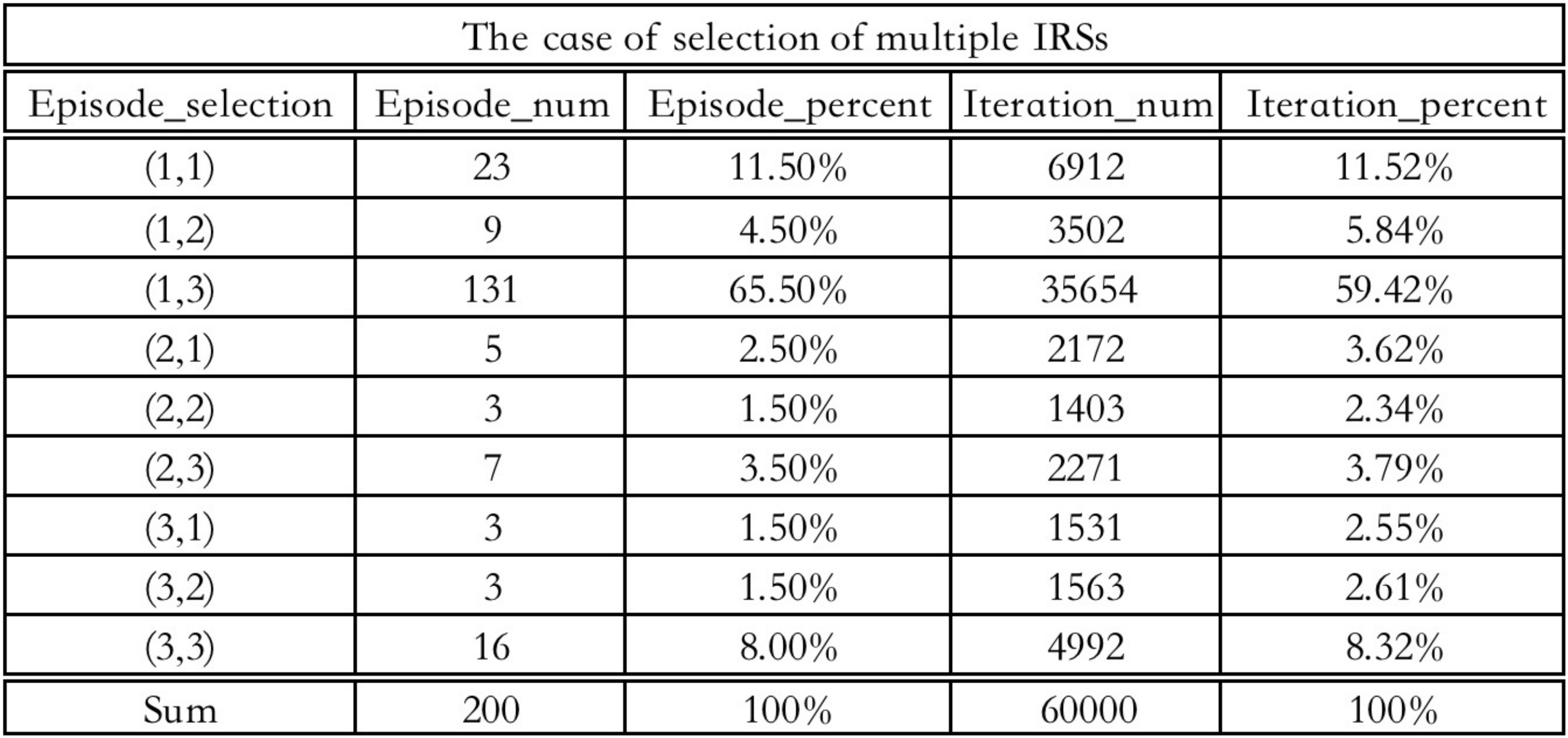} \\
      \end{minipage}}
      \subfigure[The table of IRS selection situation with the offset of $\rm Eve_{3}$.]{
        \begin{minipage}{8cm}
          \includegraphics[width=\textwidth]{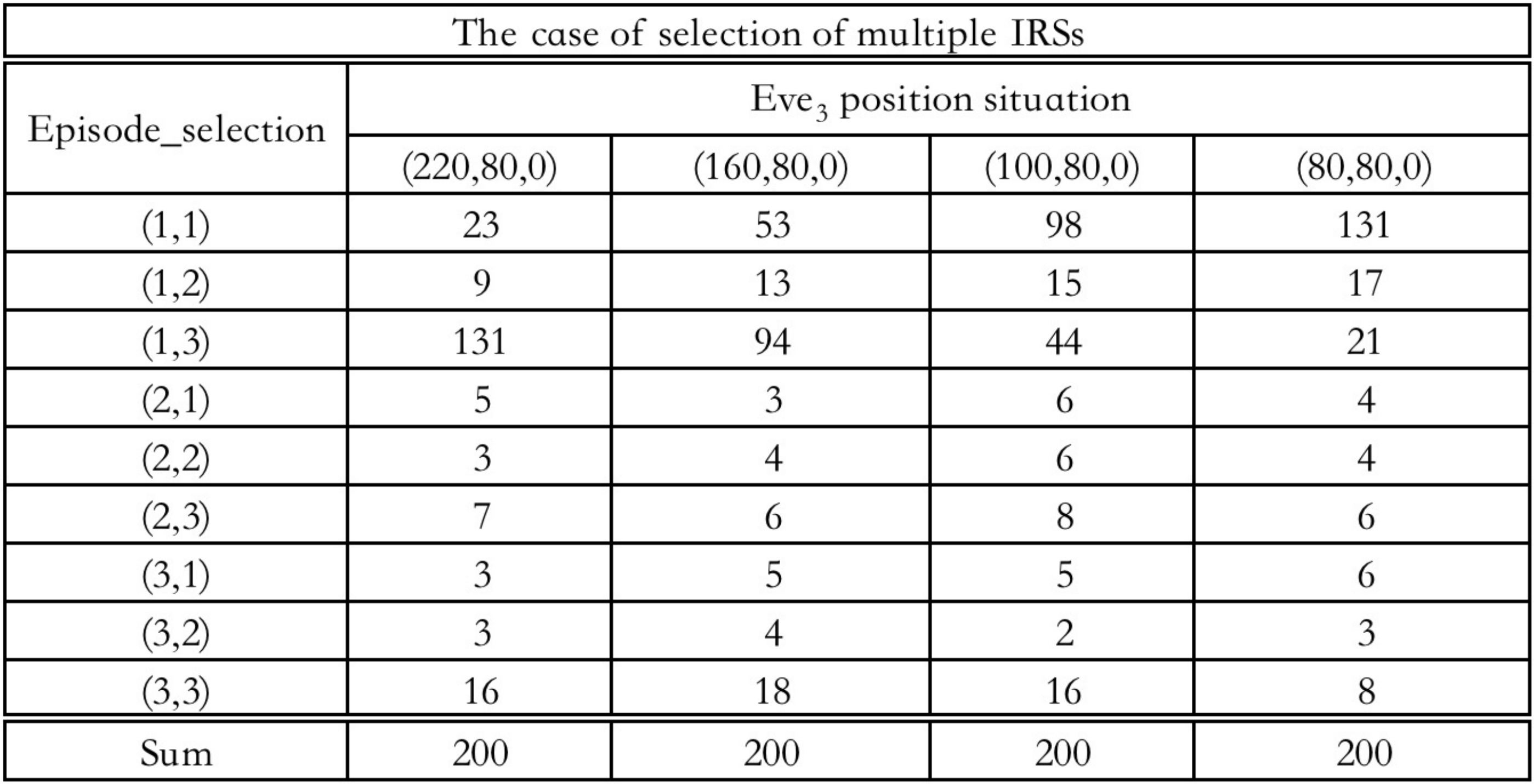} \\
        \end{minipage}}
    \caption{The detailed settings of our system model and IRS selection situation.} 
    \vspace{-0.6cm}
\end{figure}
Fig. 4(a) shows the detailed setting of our system model, including the location relationships of the PBS, SBS, SUs, PUs, Eves and IRSs.
Fig. 4(b) shows the statistics of IRS pairings after reaching convergence over two hundred cycles and sixty thousand iterations of training.
The IRS pairing array of each selection is denoted as ($z_{su_1}$,$z_{su_2}$), where the index of the array indicates the $n$-th SU and the value represents the number of times that the IRS pairing array is selected. 
From Fig. 4(a) and Fig. 4(b), it can be observed that $\rm SU_{1}$ tends to select the closest $\rm IRS_{1}$ for secure transmission. 
Thus, the conventional paradigm that SUs tend to choose the nearest IRS for pairing for transmission enhancement is confirmed.
However, $\rm SU_{2}$ tends to select the second-farthest $\rm IRS_{3}$ for secure transmission, not the closest $\rm IRS_{2}$.
The reason behind it is that the presence of eavesdroppers $\rm Eve_{2}$ is in closer proximity to $\rm IRS_{2}$ than $\rm SU_{2}$, making it a potential threat of severe eavesdropping. 
Thus, with the secure transmission in view, SUs may not choose the closest IRS, and further consideration needs to be given to the relative locations of Eves and SUs with respect to IRSs, for IRS-SU pairings.
To further investigate the effect of relative location on IRS pairings, the IRS paring situation can be shown in Fig. 4(c), where $\rm Eve_{3}$ moves from $\left(220, 80, 0\right)$ to $\left(80, 80, 0\right)$ in the $xy$-plane along the negative direction of the $x$-axis. 
From Fig. 4(c), it can be seen that as $\rm Eve_{3}$ moves closer to $\rm IRS_{3}$, $\rm SU_{2}$ gradually stops paring with $\rm IRS_{3}$ for secure transmission and tends to pair with $\rm IRS_{1}$, which is farther away than $\rm IRS_{2}$ and $\rm IRS_{3}$. 
This is mainly due to the fact that the eavesdropping of $\rm Eve_{3}$ for $\rm SU_{2}$ is enhanced as $\rm Eve_{3}$ moves closer to $\rm IRS_{3}$ and exceeds the eavesdropping rate of $\rm Eve_{2}$ on $\rm SU_{2}$ when $\rm IRS_{3}$ is used for $\rm SU_{2}$.
To summarize, the inappropriate deployment of the IRSs can reduce the security performance when Eves are present, and the arrangement of IRSs deserves further consideration.

\begin{figure}
  \centering
  \includegraphics[scale=0.55]{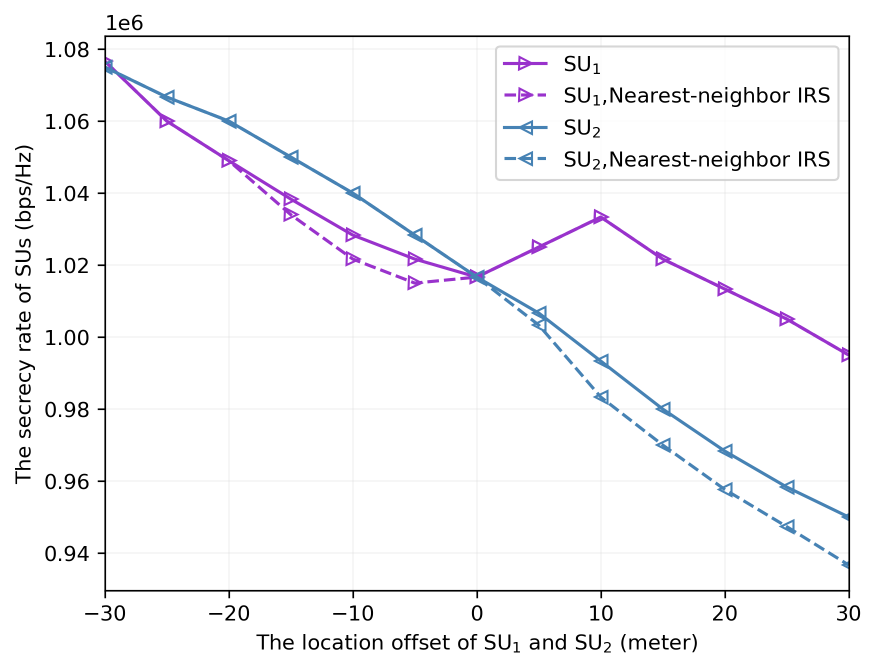}
  \vspace{-0.2cm}
  \caption{The secrecy rate versus the location distance of $\rm SU_{1}$ and $\rm SU_{2}$.}
  \vspace{-0.45cm}
\end{figure}
Fig. 5 shows the secrecy rate of the secondary network with fixed multi-IRS versus the location variation of the SUs, where $\rm SU_{1}$ is moved along the $y$-axis direction, and $\rm SU_{2}$ is moved along the $x$-axis direction. 
The nearest-neighbor scheme is introduced where the closest IRS is always selected.
Based on the numerical result, each inflection of line segments and large change in line slopes occur with IRS pairing variation or strongest eavesdropper variation.
As shown in Fig. 5, the $\rm SU_{2}$ paired with the $\rm IRS_{3}$, obtains an average $2.13\times10^3 \mathrm{~bps/Hz/m}$ secrecy loss when closing to the $\rm Eve_{2}$ during the movement.
The turning point of $\rm SU_{1}$ in offset 10 $\rm m$ represents a shift in IRS pairing from $\rm IRS_{3}$ to $\rm IRS_{1}$ for better secure transmission performance.
An $1.95\times10^3 \mathrm{~bps/Hz/m}$ secure performance loss can be seen as $\rm SU_{1}$ keeps away from the $\rm IRS_{3}$ and approaches the $\rm IRS_{1}$ and $\rm Eve_{1}$. 
The IRS paring change for $\rm SU_{1}$ achieves an improvement of $1.85\times10^3 \mathrm{~bps/Hz/m}$, from offset 0 $\rm m$ to 10 $\rm m$. 
As the angle difference from $\rm IRS_{1}$ to $\rm SU_{1}$ and $\rm Eve_{1}$ becomes slim, from offset 10 $\rm m$ to 30 $\rm m$, an average $1.9\times10^3 \mathrm{~bps/Hz/m}$ secure performance loss is observed.
Furthermore, compared to choosing the closest IRS, the experimental result in Fig. 5 further reveals that our proposed scheme can find feasible solutions for SUs in different locations, performing robustness and generalization ability.

\vspace{-0.4cm}
\subsection{Multi-IRS Assisted Secure Transmission}
\begin{figure*}
  \centering
  \includegraphics[scale=0.6]{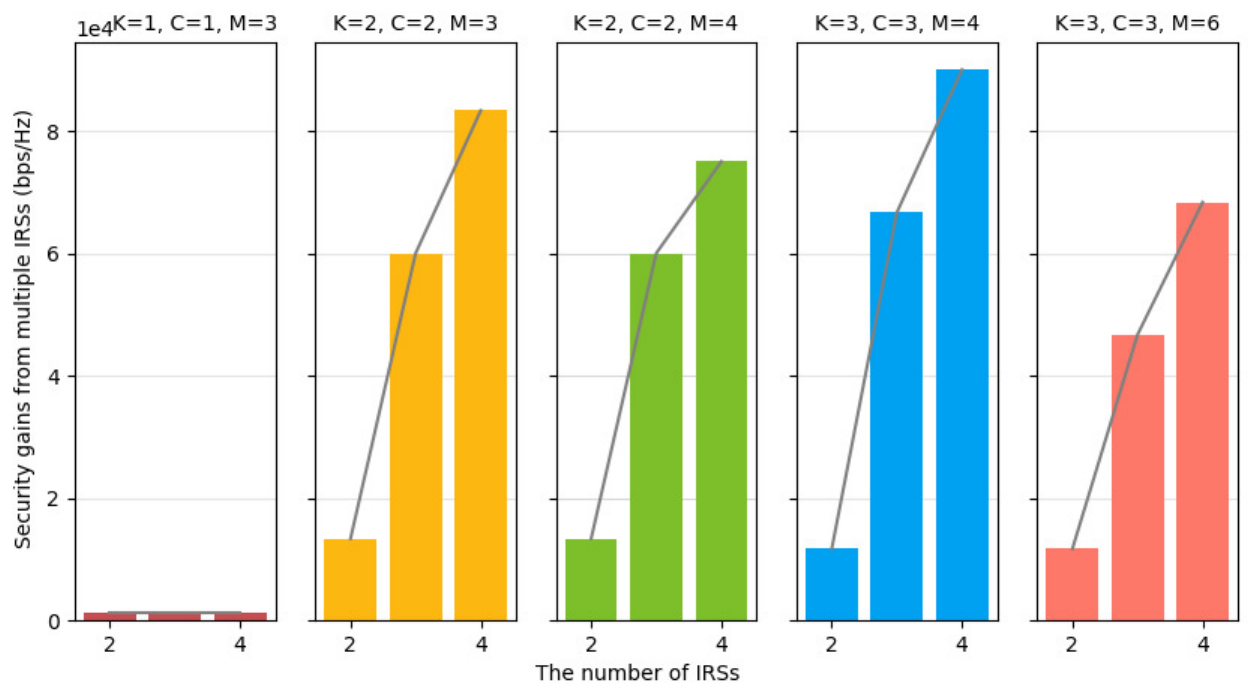}
  \vspace{-0.3cm}
  \caption{The security gains from multiple IRSs over a single IRS versus the number of multiple IRSs.}
  \vspace{-0.3cm}
\end{figure*}
Fig. 6 shows the security gains versus the number of IRSs. 
For better comparison, additional SU, PU, Eve and IRS are added to the proposed scenario. 
Specifically, a fourth IRS is deployed at $\left(120, 90, 20\right)$, a third SU is deployed at $\left(110, 100, 0\right)$, and a third PU is deployed at $\left(250, 100, 0\right)$.
The fourth, fifth and sixth Eves are respectively deployed at $\left(130, 140, 0\right)$, $\left(120, 20, 0\right)$ and $\left(20, 170, 0\right)$.
As can be observed in Fig. 6, with $K = 1$, $C = 1$, and $M = 3$, the secrecy rate of SUs is hardly improved with the number of IRSs. 
This is because $\rm IRS_{1}$ is the optimal paring for $\rm SU_{1}$, and the additional IRSs make a little contribution.
When the third IRS is applied, the security rate is significantly improved due to the ability of $\rm IRS_{3}$ to provide a better secure transmission for $\rm SU_{2}$, which is consistent with the results in Fig. 4.
With an increasing number of users, the deployment of multiple IRSs is fully utilized and the achievable secrecy rate is improved. 
When $K = 2$ and $C = 2$, comparing the number of the Eves from $M = 3$ to $M = 4$, the secrecy rate improvement witnesses a decline due to a stronger eavesdropping ability that the additional Eves have.
It shows that our proposed scheme can still take full advantage of the multiple IRSs, finding the feasible suboptimal solutions with the increasing number of the Eves.
Moreover, It can be seen that the increase in the number of IRSs only helps the users who are paired with the newly added IRS, which further illustrates the importance of the deployment of IRSs in enhancing the secrecy rate.

\begin{figure}[H]
  \centering
  \includegraphics[scale=0.55]{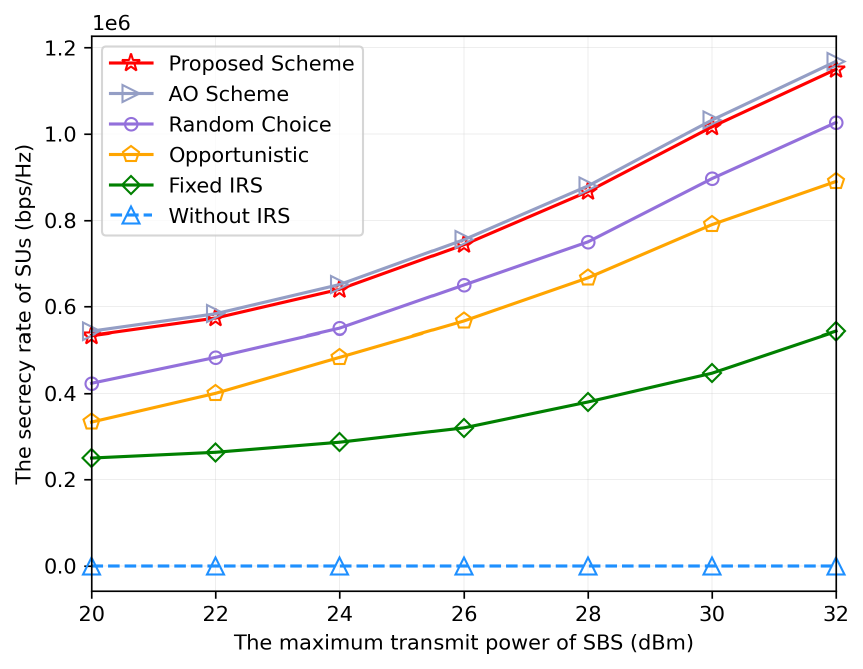}
  \vspace{-0.3cm}
  \caption{The secrecy rate versus the SBS maximum transmit power $TP_s$ under different schemes.}
\end{figure}
Fig. 7 compares the secrecy rate of our proposed scheme with the benchmark schemes versus the maximum transmit power $TP_s$. 
As shown in Fig. 7, our proposed scheme outperforms benchmark schemes, and the achievable secrecy rate of SUs with IRS assistance increases with the transmit power of the SBS. 
Due to the computational complexity and CPU time comparison shown in TABLE I, the slight secure performance loss of our proposed intelligent scheme compared to the AO algorithm can be tolerated.
At the same time, the spectrum-aware enhancement scheme performs better than the opportunistic access scheme in terms of the secrecy rate in the sense that our proposed sensing-enhanced spectrum sharing scheme achieves better spectrum utilization efficiency than the opportunistic access scheme. 
It is also noteworthy that the secure transmission is kept at 0 in the scheme without IRS augmentation, 
which indicates that the eavesdropping rate of Eves can be larger than the transmission rate of SUs, demonstrating that the application of IRSs benefits secure transmission.
  
\begin{figure}
  \centering
  \includegraphics[scale=0.55]{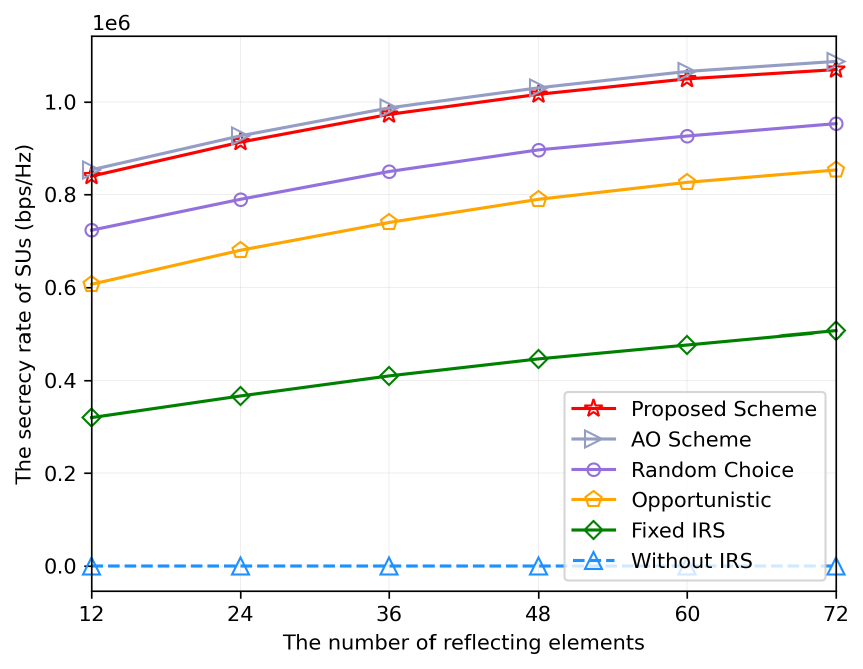}
  \vspace{-0.3cm}
  \caption{The secrecy rate versus the number of reflection elements of each IRS.}
  \vspace{-0.2cm}
\end{figure}
Fig. 8 compares the secrecy rate of our proposed scheme with the benchmark schemes versus the number of passive reflection elements of each IRS. 
From Fig. 8, the sum secrecy rate of the IRS-assisted secure transmission schemes increases with the number of IRS reflection elements. 
This is because the reflection array elements of multiple IRSs are able to enhance the signal received by SUs and attenuate the signal received by Eves. 
Moreover, it can be seen in Fig. 8 that our proposed scheme outperforms benchmark schemes and obtains a better increment in the secrecy rate with the number of IRS reflection elements.
This is because our proposed intelligent scheme can steadily select the most efficient IRS and dynamically adjust the reflective array of the selected IRSs.
Furthermore, as the number of IRS reflective elements increases, the performance loss of our proposed intelligent scheme is still fairly slim, which is acceptable considering the computing complexity advantage.

\vspace{-0.3cm}
\section{Conclusion}
\vspace{-0.1cm}
A multi-IRS-assisted sensing-enhanced wideband spectrum sharing network with the presence of eavesdropping was studied in this paper. 
The sum secrecy rate of the secondary network was maximized by jointly optimizing subchannel assignment, IRS pairing, transmit beamforming of the SBS, reflection coefficients of IRSs, and the sensing time while taking the QoS requirements of PUs into account. 
To address the challenging non-convex optimization problem with coupled variables and integer programming, an intelligent resource allocation scheme was designed to overcome the difficulty of the mixed and high-dimensional action space issue by exploiting the H2DS framework.
An AO-based traditional mathematical method is considered to highlight the computational efficiency of our proposed intelligent algorithm.
Simulation results demonstrated that our proposed intelligent scheme outperformed benchmark schemes in terms of the secrecy rate.  
Moreover, it was also shown that the inappropriate deployment of IRSs could reduce the security performance when Eves were present, and the arrangement of IRSs deserves further consideration. 

\vspace{-0.2cm}
\section*{Appendix A}
\vspace{-0.1cm}
\section*{The processing of the AO approach}
\vspace{-0.1cm}
Given the substantial count of reflective array elements of multiple IRSs necessitating optimization, and acknowledging that the alternating iteration algorithm solely requires the storage of a singular variable from the subproblem in each iteration, the alternating iteration algorithm is employed to tackle the large-scale intricate non-convex optimization problems with mixed integer programming.
Initially, the transmit beamforming of the SBS and the subchannel allocation are jointly optimized based on the given IRS pairing, IRS reflection coefficients, and the sensing time. Subsequently, we focus on optimizing IRS pairing, leveraging the derived transmit beamforming and subchannel allocation. Moving forward, the IRS reflection coefficients are optimized, informed by parameters obtained from the preceding optimization stages. Finally, a one-dimensional search algorithm is employed to determine the optimal sensing time.

\vspace{-0.2cm}
\subsection{Joint Optimization of Transmit Beamforming Allocation and Subchannel Allocation}
To address the efficient resolution of $\mathbf{P1}$, the indicator variable for the subchannel assignment, $\rho_{n, k}$, is treated as a continuous variable $\rho_{n, k} \in [0, 1]$. 
Furthermore, auxiliary variables $x_{k,c}^0$ and $x_{k,c}^1$ are introduced, effectively decoupling the relationship between $\xi_{k, c}$ and $\mathbf{f}_{k, c}^{s,j}$, $j \in \left\{0, 1\right\}$,
where $\mathbf{f}_{k,c}^{s,0}$ denotes the transmit beamforming of the SBS when the subchannel is sensed to be idle, $\mathbf{f}_{k,c}^{s,j}$ denotes the transmit beamforming of the SBS when the subchannel is sensed to be active, and $x_{k,c}^{j} = \xi_{k,c} \left| \mathbf{f}_{k,c}^{s,j} \right|^2$.
Consequently, for the given IRS pairing $\zeta^*_{k,z}$, the IRS reflection coefficients $\Phi^*_{z}$ and the sensing time $\tau^*$, 
the comprehensive optimization problem of the transmit beamforming and subchannel allocation is given as
\begin{subequations}
  \begin{align}
  \mathbf{P2}: &\underset{\xi_{k,c}, x_{k,c}^{0}, x_{k,c}^{1}}{\operatorname{max}} \sum_{c=1}^C \sum_{k=1}^K \xi_{k,c} r^{\sec}_{k,c} \\ 
  \text{ s.t. }  
  &\mathrm{C} 1: \frac{T-\tau^*}{T} Y_c \sum_{k=1}^{K} \sum_{i=0}^{1} P^{j1} x_{k,c}^{j} \leq I_{th}^s, \forall c \in \mathcal{C},\\
  &\mathrm{C} 3: \frac{T-\tau^*}{T} \sum_{k=1}^K\sum_{c=1}^C \sum_{j=0}^{1} P^{0j} x_{k,c}^{0} + P^{1j} x_{k,c}^{1}\leq P_{th}^s,\\
  &\mathrm{C} 6: \xi_{k,c} \in [0,1],\sum_{k=1}^{K} \xi_{k,c}=1, \forall c \in\mathcal{C}, \forall k \in\mathcal{K},\\
  &\mathrm{C} 2, 
  \end{align}
\end{subequations}
where $Y_c$ denotes the channel power gain between the SBS and the PU over the $c$-th subchannel, which can be denoted by
$Y_c = \left|\mathbf{h}_{s, d}^T+\mathbf{g}_{{z_k}, d}^T (\sum_{z_k=1}^{Z}\xi_{k,z_k}^* \mathbf{\Phi}_{z_k}^*) \mathbf{G}_{z_k}\right|^2$.
The $I_{th}^s$ denotes the maximum interference tolerance of the PUs.
The $r^{\sec}_{k,c}$ denotes the secure transmission rate, which is represented by
$r^{\sec}_{k,c} = \frac{T-\tau^*}{T} \sum_{j=0}^{1} \operatorname{P}^{0j}\left(\Re_{k}^{0j}-\underset{\forall m}{\operatorname{max}}\Re_{m,k}^{0j}\right)^{+}+\operatorname{P}^{1j}\left(\Re_{k}^{1j}-\underset{\forall m}{\operatorname{max}}\Re_{m,k}^{1j}\right)^{+}$,
where 
\begin{subequations}
  \begin{align}
&\Re_{k}^{0j} = \log_2 \left(1+\frac{\left|\left(\mathbf{h}_{s, k}^T+\mathbf{g}_{{z}, k}^T (\sum_{z=1}^{Z}\xi_{k,z}^* \mathbf{\Phi}_{z}^*) \mathbf{G}_{s, R_{z}}\right) \right|^2 x_{k,c}^{j}}{\sigma_{k}^2}\right),\\
&\Re_{k}^{1j} = \log_2 \nonumber \\
&\left(1+\frac{\left|\left(\mathbf{h}_{s, k}^T+\mathbf{g}_{{z}, k}^T (\sum_{z=1}^{Z}\xi_{k,z}^* \mathbf{\Phi}_{z}^*) \mathbf{G}_{s, R_{z}}\right) \right|^2 x_{k,c}^{j}}{\sum\limits_{d=1}^D \delta_{d,c}\left|\left(\mathbf{h}_{p, k}^T+\mathbf{g}_{{z_d}, k}^T \mathbf{\Phi}^*_{z_d} \mathbf{H}_{p, R_{z_d}}\right) {\mathbf{f}}_{d,c}^p\right|^2+\sigma_{k}^2}\right),\\
&\Re_{m,k}^{0j} = \log_2 \nonumber \\
&\left(1+\frac{\left|\left(\mathbf{h}_{s, m}^T+\mathbf{g}_{{z_k}, m}^T (\sum_{z_k=1}^{Z}\xi_{k,z_k}^* \mathbf{\Phi}_{z_k}^*) \mathbf{G}_{z_k}\right) \right|^2 x_{k,c}^{j}}{\sigma_{m}^2}\right),\\
&\Re_{m,k}^{1j} = \log_2 \nonumber \\
&\left(1+\frac{\left|\left(\mathbf{h}_{s, m}^T+\mathbf{g}_{{z_k}, m}^T (\sum_{z_k=1}^{Z}\xi_{k,z_k}^* \mathbf{\Phi}_{z_k}^*) \mathbf{G}_{z_k}\right) \right|^2 x_{k,c}^{j}}{\sum\limits_{d=1}^D \delta_{d,c}\left|\left(\mathbf{h}_{p, m}^T+\mathbf{g}_{{z_d}, m}^T \mathbf{\Phi}^*_{z_d} \mathbf{H}_{z_d}\right) \mathbf{f}_{d,c}^p\right|^2+\sigma_{m}^2}\right).
\end{align}
\end{subequations}

It is apparent that $\mathbf{P2}$ is convex. Therefore, the Lagrangian approach is applied. 
The Lagrangian of $\mathbf{P2}$ is presented by (35).
\begin{figure*}
\begin{equation}
  \begin{aligned}
  \mathcal{L}\left(\xi_{k,c}, x_{k,c}^{0}, x_{k,c}^{1}\right)& =\sum_{c=1}^C \sum_{k=1}^K \xi_{k,c} r^{\sec}_{k,c} +\sum_{c=1}^{C}\omega_c\left[I_{th}^s-\frac{T-\tau^*}{T} Y_c \sum_{k=1}^{K} \sum_{j=0}^{1} P^{j1} x_{k,c}^{j}\right]\\
  & +\varsigma\left(P_{th}^s -\frac{T-\tau^*}{T} \sum_{k=1}^K\sum_{c=1}^C \sum_{j=0}^{1} P^{0j} x_{k,c}^{0} + P^{1j} x_{k,c}^{1}\right)+\sum_{c=1}^C v_c\left(1-\sum_{k=1}^K \xi_{k, c}\right) .
  \end{aligned}
\end{equation}
\hrulefill
\end{figure*}

In accordance with the Karush-Kuhn-Tucker (KKT) conditions, the following relations can be obtained as
\begin{subequations}
  \begin{align}
  & \frac{\partial L}{\partial x_{k,c}^0}\left\{\begin{array}{c}
  =0, x_{k,c}^0>0 \\
  <0, x_{k,c}^0=0
  \end{array} \quad \forall k, c,\right. \\
  & \frac{\partial L}{\partial x_{k,c}^1}\left\{\begin{array}{l}
  =0, x_{k,c}^1>0 \\
  <0, x_{k,c}^1=0
  \end{array} \quad \forall k, c,\right. \\
  & \frac{\partial L}{\partial \xi_{k,c}}\left\{\begin{array}{l}
  <0, \xi_{k,c}>0 \\
  =0,0<\xi_{k,c}<1 \\
  <0, \xi_{k,c}=1
  \end{array} \quad \forall k, c .\right.
  \end{align}
\end{subequations}

Followed by (36a) and (36b), the optimal transmit beamforming, denoted as $\widetilde{\mathbf{f}}_{k,c}^{s,j}$ , can be represented by
\begin{subequations}
  \begin{align}
  & A_k=\left|\mathbf{h}_{s, k}^T+\mathbf{g}_{{z}, k}^T (\sum_{z=1}^{Z}\xi_{k,z}^* \mathbf{\Phi}_{z}^*) \mathbf{G}_{s, R_{z}}\right|^2, \\
  & B_{m,k}=\left|\mathbf{h}_{s, m}^T+\mathbf{g}_{{z_k}, m}^T (\sum_{z_k=1}^{Z}\xi_{k,z_k}^* \mathbf{\Phi}_{z_k}^*) \mathbf{G}_{z_k}\right|^2, \\
  & z_k = \left|\mathbf{h}_{p, k}^T+\mathbf{g}_{{z_c}, k}^T \mathbf{\Phi}^*_{z_c} \mathbf{H}_{p, R_{z_c}}\right|^2,\\
  & z_m = \left|\mathbf{h}_{p, m}^T+\mathbf{g}_{{z_c}, m}^T \mathbf{\Phi}^*_{z_c} \mathbf{H}_{z_c} \right|^2,\\
  & c_{c}^j=\ln 2\left[\omega_c Y_c {\rm P}^{j,1}+\varsigma\left({\rm P}^{j0}+{\rm P}^{j1}\right)\right], \\
  & b_{k,c}^j=c_{c}^{j} \left[A_k\left(\left|{\mathbf{f}}_{c}^p\right|^2 z_k +2 \sigma_k^2\right)-B_{\widehat{m},k}\left(\left|{\mathbf{f}}_{c}^p\right|^2 z_m +2 \sigma_m^2\right) \right] \nonumber \\
  & -(A_k^2-B_{\widehat{m},k}^2)\left({\rm P}^{j0}+{\rm P}^{j1}\right), \\
  & s_{k,c}^j=\left(\sigma_k^2 c_{c}^{j}-A_k{\rm P}^{j0}\right)\left(\left|{\mathbf{f}}_{c}^p\right|^2 z_k+\sigma_k^2\right)-\left(\sigma_m^2 c_{c}^{j}-B_{\widehat{m},k}{\rm P}^{j0}\right) \nonumber \\
  & \left(\left|{\mathbf{f}}_{c}^p\right|^2 z_m+\sigma_m^2\right)-{\rm P}^{j1} (A_k\sigma_k^2-B_{\widehat{m},k}\sigma_m^2),\\
  &\widetilde{\mathbf{f}}_{k,c}^{s,j}=\left[\frac{\left(-b_{k,c}^j \pm \sqrt{(b_{k,c}^j)^2-4 c_{c}^j s_{k,c}^j}\right)}{2 (A_k^2-B_{\widehat{m},k}^2) c_{c}^j}\right]^{+},
  \end{align}
\end{subequations}
where $\varsigma \leq 0$ and $\omega_c \leq 0$ are the dual variables associated with constraints $\mathrm{C} 3$ and $\mathrm{C} 6$, respectively.
The $\widehat{m}$ denotes the eavesdropper with the largest rate of eavesdropping.

The optimal subchannel allocation, denoted by $\widetilde{\xi}_{k,c}$, can be obtained by \textbf{Theorem 1}.

\textbf{Theorem 1}: The optimal subchannel allocation for the joint optimization of the transmit beamforming and subchannel allocation is obtained as
\begin{subequations}
  \begin{align}
  &i(c)=\arg \max _{k \in \mathcal{K}} H_{k, c}, \widetilde{\xi}_{i(c),c}=1, \widetilde{\xi}_{k, c}=0, \forall k \neq i(c),\\
  &H_{k, c}=\left(\frac{T-\tau^*}{T}\right)\nonumber \\
  &\left[{\rm P}^{0,j}\left(\Re_{k}^{0,j}-H_{k, c}^{0,j}\right)+{\rm P}^{1,j}\left(\Re_{k}^{1,j}-H_{k, c}^{1,j}\right)\right] .
  \end{align}
\end{subequations}
Based on \textbf{Theorem 1}, the $k$-th subchannel is allocated to the SUs with the largest $H_{k, c}, k \in \mathcal{K}$.

\textbf{Proof:} In light of the KKT conditions, it follows
\begin{equation}
\frac{\partial \mathcal{L}\left(\xi_{k,c}, x_{k,c}^{0}, x_{k,c}^{1}\right)}{\partial \xi_{k,c}}\left\{\begin{array}{l}
<0, \xi_{k,c}>0, \\
=0,0<\xi_{k,c}<1, \quad \forall k, c . \\
<0, \xi_{k,c}=1 .
\end{array}\right.
\end{equation}
The optimal subchannel allocation is represented by
\begin{subequations}
  \begin{align}
& \widetilde{\xi}_{k,c}=\left\{\begin{array}{l}
0, H_{k, c}<v_c, \\
1, H_{k, c}>v_c,
\end{array}\right. \\
& H_{k, c}=\left(\frac{T-\tau^*}{T}\right)\left[{\rm P}^{j, 0}\left(r_{k, c}^{j, 0}-H_{k, c}^{j, 0}\right)+{\rm P}^{j, 1}\left(r_{k, c}^{j, 1}-H_{k, c}^{j, 1}\right)\right], \\
& H_{k, c}^{j, 0}=\frac{A_k\left|\mathbf{f}_{k,c}^{s,j}\right|^{2}}{\ln 2\left(\sigma_k^2+A_k\left|\mathbf{f}_{k, c}^{s}\right|^2\right)} - \frac{B_{\widehat{m}}\left|\mathbf{f}_{k,c}^{s,j}\right|^{2}}{\ln 2\left(\sigma_m^2+B_{\widehat{m}}\left|\mathbf{f}_{k, c}^{s}\right|^2\right)}, \\
& H_{k, c}^{j, 1}=\frac{A_k\left|\mathbf{f}_{k,c}^{s,j}\right|^{2}}{\ln 2\left[\left(\left|\mathbf{\mathbf{f}}_{c}^{p}\right|^2 z_k+\sigma_k^2\right)+A_k\left|\mathbf{f}_{k, c}^{s,j}\right|^2\right]}\nonumber \\
&-\frac{B_{\widehat{m}}\left|\mathbf{f}_{k,c}^{s,j}\right|^{2}}{\ln 2\left[\left(\left|\mathbf{\mathbf{f}}_{c}^{p}\right|^2 z_m+\sigma_m^2\right)+B_{\widehat{m}}\left|\mathbf{f}_{k, c}^{s,j}\right|^2\right]},
  \end{align}
\end{subequations}
where $H_{k, c}$ is the subchannel subchannel indicator, which stands as a pivotal determinant in the pursuit of the optimal channel allocation and transmit beamforming. 
The indicator encapsulates the delicate balance between the secure rate and interference. 
The proof for \textbf{Theorem 1} is thereby concluded.

\vspace{-0.5cm}
\subsection{Optimization of IRS Pairngs}
Given the IRS reflection coefficients $\Phi^*_{z}$, the sensing time $\tau^*$, the optimal transmit beamforming allocation $\widetilde{\mathbf{f}}_{k,c}$ and the optimal subchannel allocation $\widetilde{\xi}_{k,c}$, 
the optimization of IRS pairing can be formulated as
\begin{subequations}
  \begin{align}
  \mathbf{P3}: &\underset{\zeta_{k,z}}{\operatorname{max}} \sum_{c=1}^C \sum_{k=1}^K \xi_{k,c} r^{\sec}_{k,c} \\ 
  \text{ s.t. }  
  &\mathrm{C} 1: \frac{T-\tau^*}{T} \sum_{z=1}^{Z} \zeta_{k,z}\left(\mathbf{h}_{s,d}^T+\mathbf{g}_{{z_k}, d}^T \boldsymbol{\Phi}^*_{z_k} \mathbf{G}_{z_k}\right) \nonumber \\
  &\sum_{k=1}^{K} \sum_{j=0}^{1} P^{j1} \widetilde{x}_{k,c}^{j} \leq I_{th}^s, \forall c \in \mathcal{C},\\
  &\mathrm{C} 7: \zeta_{k,z} \in [0,1], \sum_{z=1}^{Z} \zeta_{k,z}=1, \forall z \in \mathcal{Z},\forall k \in \mathcal{K},\\
  &\mathrm{C4}, \mathrm{C5},
  \end{align}
\end{subequations}
where $\widetilde{x}_{k,c}^{i} = \widetilde{\xi}_{k,c} \left| \widetilde{\mathbf{f}}_{k,c}^{s,i} \right|^2$.
It is obvious that $\mathbf{P3}$ is convex. Hence, similar to the solution of the subchannel assignment in \textbf{P2}, the suboptimal IRS pairngs are obtained.

\vspace{-0.5cm}
\subsection{Optimization of IRS Reflection Coefficients}
Given the sensing time $\tau^*$, the optimal IRS pairing $\widetilde{\zeta}_{k,z}$, the optimal transmit beamforming allocation $\widetilde{\mathbf{f}}_{k,c}$ and the optimal subchannel allocation $\widetilde{\xi}_{k,c}$, 
the optimization of IRS reflection coefficients can be formulated as
\begin{subequations}
  \begin{align}
  \mathbf{P4}: &\underset{\mathbf{\Theta}}{\operatorname{max}} \sum_{c=1}^C \sum_{k=1}^K \xi_{k,c} r^{\sec}_{k,c} \\ 
  \text{ s.t. }  
  &\left|\phi_{z,n}\right| \leq 1, \forall m,
  \end{align}
\end{subequations}
It is evident that $\mathbf{P3}$ is non-convex owing to its non-convex objective function. 
As a result, the expression for the combined channel power gain of the link connecting the CBS to the $n$-th SU is reformulated as
\begin{equation}
  \left|\mathbf{h}_{s, k}^T+\mathbf{g}_{{z}, k}^T \mathbf{\Phi}_{z} \mathbf{G}_{s, R_{z}}\right|^2 =\left|\mathbf{h}_{s, k}^T+\boldsymbol{\vartheta}_z^H \boldsymbol{\phi}_z\right|^2,
\end{equation}
where $\boldsymbol{\vartheta}^H_z=\mathbf{g}_{z,k}^H \operatorname{diag}\left(\mathbf{G}_{z,n}\right)$. 
To obtain the optimal solution, the successive convex approximation (SCA) technique is adopted to address $\mathbf{P3}$. 
Let $\boldsymbol{\mu}, \boldsymbol{\lambda}$ and $\boldsymbol{\kappa}$ denote slack variables, where $\mu_{k,z}=\operatorname{Re}\left(\mathbf{h}_{s, k}^T+\boldsymbol{\vartheta}_z^H \boldsymbol{\phi}_z\right), \forall k, 
\lambda_{k,z}=\operatorname{Im}\left(\mathbf{h}_{s, k}^T+\boldsymbol{\vartheta}_z^H \boldsymbol{\phi}_z\right), \forall k$, 
and $\boldsymbol{\kappa}_{k,z}=\left|\mathbf{h}_{s, k}^T+\boldsymbol{\vartheta}_z^H \boldsymbol{\phi}_z\right|^2, \forall n$ respectively. 
Hence, the equivalent problem of $\mathbf{P4}$ can be formulated as
\begin{subequations}
  \begin{align}
& \mathbf{P_{4.1}}: \max _{\boldsymbol{\phi}, \boldsymbol{\mu}, \boldsymbol{\lambda}, \boldsymbol{\kappa}} \sum_{c=1}^C \sum_{k=1}^K \xi_{k,c} r^{\sec}_{k,c}\left(\kappa_{k,z}\right) \\
\text { s.t. } &\left|\phi_{z,n}\right| \leq 1, \forall n, \\
& \mu_{k,z}=\operatorname{Re}\left(\mathbf{h}_{s, k}^T+\boldsymbol{\vartheta}_z^H \boldsymbol{\phi}_z\right), \forall k, \\
& \lambda_{k,z}=\operatorname{Im}\left(\mathbf{h}_{s, k}^T+\boldsymbol{\vartheta}_z^H \boldsymbol{\phi}_z\right), \forall k, \\
& \kappa_{k,z} \leq \mu_{k,z}^2+\lambda_{k,z}^2, \forall k.
\end{align}
\end{subequations}

Let $F\left(\mu_{k,z}, \lambda_{k,z}\right) \triangleq \mu_{k,z}^2+\lambda_{k,z}^2$ denote the convex and differentiable function based on $\mu_{k,z}$ and $\lambda_{k,z}$. 
Given the local points $\left(\mu_{k,z}^*, \lambda_{k,z}^*\right)$,
the lower bound of $\mu_{k,z}^2+\lambda_{k,z}^2$ is calculated by using the first order Taylor expansion, represented by
\begin{equation}
\begin{aligned}
F\left(\mu_{k,z}, \lambda_{k,z}\right) & \geq \mu_{k,z}^{* 2}+\lambda_{k,z}^{* 2}+2 \mu_{k,z}^*\left(\mu_{k,z}-\mu_{k,z}^*\right) \\
& +2 \lambda_{k,z}^*\left(\lambda_{k,z}-\lambda_{k,z}^*\right) \triangleq \bar{F}\left(\mu_{k,z}, \lambda_{k,z}\right) .
\end{aligned}
\end{equation}
Thus, $\mathbf{P_{4.1}}$ is reformulated as
\begin{subequations}
  \begin{align}
\mathbf{P_{4.2}}: \max _{\boldsymbol{\phi}, \boldsymbol{\mu}, \boldsymbol{\lambda}, \boldsymbol{\kappa}} & \sum_{n=1}^N \sum_{c=1}^C \sum_{k=1}^K \xi_{k,c} r^{\sec}_{k,c}\left(\kappa_{k,z}\right) \\
\text { s.t. } & (15 b)-(15 d), \\
& \kappa_{k,z} \leq \bar{F}\left(\mu_{k,z}, \lambda_{k,z}\right), \forall n .
\end{align}
\end{subequations}
Problem $\mathbf{P_{4.2}}$ is convex, rendering it amenable to effective resolution by using CVX programming tool \cite{grant2017cvx}. Consequently, the approximated solution for $\mathbf{P4}$ can be attained through iterative updates of $\boldsymbol{\mu}$ and $\boldsymbol{\lambda}$ based on the optimal solution derived from $\mathbf{P_{3.2}}$.

Ultimately, the one-dimensional search approach is applied to obtain the optimal sensing time, similar to the approaches outlined in \cite{7946279}.
With the aforementioned steps, the alternative optimization algorithm solve the $\mathbf{P1}$.

\bibliography{IEEEabrv,mybibfile}

\end{document}